\documentclass[onecolumn,linenumbers]{revtex4}
\usepackage{graphicx}
\usepackage{graphicx}
\usepackage{dcolumn}
\usepackage{bm}
\usepackage[latin1]{inputenc}
\usepackage{amsmath}
\usepackage{amsfonts}
\usepackage{subfigure}
\usepackage{amssymb}
\usepackage{mathrsfs}
\RequirePackage[colorlinks,citecolor=blue,urlcolor=blue,linkcolor=magenta]{hyperref}
\begin{document}
\title{Late time accelerated scaling attractors in DGP (Dvali-Gabadadze-Porrati) braneworld}
\smallskip

\author{\bf~Jibitesh~Dutta $^{1}$\footnote{jdutta29@gmail.com,~jibitesh@nehu.ac.in},
 Wompherdeiki Khyllep  $^2$\footnote{sjwomkhyllep@gmail.com},  Erickson Syiemlieh $^2$\footnote{eric.syiem@gmail.com}}

\smallskip

\affiliation{$^{1}$Mathematics Division, Department of Basic
Sciences and Social Sciences,~ North Eastern Hill University,~NEHU
Campus, Shillong - 793022 , Meghalaya ( INDIA )}

\affiliation{$^2$ Department of Mathematics, St. Anthony's College, Shillong - 793001, Meghalaya ( INDIA )}

\date{\today}

\begin{abstract}
\noindent In the evolution of late universe, the main source of matter are Dark energy and Dark matter. They are indirectly detected only through their gravitational manifestations.
 So the possibility of interaction with each other without violating observational restrictions is not ruled out. With this motivation, we investigate the dynamics of DGP braneworld
 where source of dark energy is a scalar field and it interacts with matter source. Since observation favours phantom case more, we have also studied the dynamics of interacting phantom
 scalar field. In non interacting DGP braneworld there are no late time accelerated scaling attractors and hence cannot alleviate Coincidence problem. In this paper, we shall show that it
  is possible to get late time accelerated scaling solutions.  The phase space is studied by taking two categories of potentials (Exponential and Non exponential functions). The stability of
   critical points are examined by taking two specific interactions. The first interaction gives late time accelerated scaling solution for phantom field only under exponential potential, while
    for second interaction we do not get any scaling solution. Furthermore, we have shown that this  scaling solution is also classically stable.
\end{abstract}
\maketitle

\section{Introduction}\label{DGP I:Sec1:Intro}
\noindent The fact that our universe is currently undergoing an accelerated expansion has been confirmed by many observations since last fifteen years \cite{Perlmutter, Spergel,Riess,Ade}.
In standard cosmology, this accelerated expansion can be explained by dark energy(DE). It is an exotic entity with negative pressure.
One of the most simple contender for DE is the time-independent cosmological constant ($\Lambda$) whose equation of state (EOS) $\omega$ is equal to $-1$.
But this suffer from well known Coincidence Problem - why DE and Dark matter (DM) energy densities are of same order at present even though they evolve at a highly different red-shift \cite{Carroll,Peebles}. The cosmological constant problem can also be alleviated by modelling DE with a scalar field whose equation of state varies dynamically. Scalar fields play a crucial role in cosmology because they are simple and able to generate meaningful dynamics. Canonical scalar fields can be used to model dark energy and inflation\cite{Lyth,Matos}. Dynamical scalar field models such as Quintessence ($-1<\omega<-\frac{1}{3}$)\cite{Linder}, K-essence \cite{Armendariz}, Phantom fields $(\omega<-1)$ \cite{Caldwell} etc were proposed as possible candidates for DE.  For a review on different cosmological dark energy models see \cite{cope,Bamba}. These models have some merits over the cosmological constant problem. Furthermore, these models yield observed values of $\omega$ and can mimic cosmological constant at present epoch.
\par Generally, in cosmological models where scalar fields are used to describe DE, the background matter does not interact with scalar field. But there is no principle of physics by which this interaction of DE and DM can be ruled out. We may get similar energy density in the dark sector if there is an interaction of DE and DM. So the main motivation for taking interaction between DE and DM is to alleviate the coincidence problem. Since the nature of DE  and DM is still unknown, so currently there is no specific form interaction. Therefore, any interaction considered is phenomenological one, even though some may have better justification over the others.
\par The first interaction between scalar field and matter and its various ramifications are studied in \cite{Kalara,Wetterich}. The main advantage of interaction of quintessence and DM is that scaling solutions can lead to late time acceleration \cite{Wetterich,Amendola}. This cannot be obtained without considering interaction. Furthermore, attempts have been made to study the dynamics of interacting phantom fields and DM\cite{Guo,Nojiri}.
\par The recent accelerated expansion of the universe can also be explained by theories of extra dimensions. Braneworld scenario is an important theory of extra dimension inspired by string theory. In this set up, our observable universe is a hyperspace (called brane) embedded in a higher dimensional space-time known as bulk. Braneworld models correct standard cosmology in a noble way and solve many outstanding problems of cosmology. Moreover, it has some important differences from standard cosmology and one can see the standard review in \cite{Marteens}.
\par DGP braneworld (proposed by Dvali, Gabadadze and Porrati) is one of the promising theory of braneworld \cite{dgp1,dgp2}. It consists of two branches, one of which is the self-accelerating branch which does not need any DE for acceleration and the other is the normal branch which requires DE to accelerate. But the former suffers from the ghost problem while the latter is free from ghost instabilities. Usually phantom fluid is known to violate Weak Energy Condition (WEC) but in DGP (normal branch) model the phantom characteristic of violating WEC is suppressed by brane gravity effects. Another interesting feature of this brane world model is violation of Strong Energy Condition (SEC) and as a result this model accelerates \cite{JDutta1}.
\par Moreover, in the normal branch of DGP model, the generalized second law of thermodynamics (GSLT) is satisfied at both apparent
and event horizon (with some reasonable restrictions) and as a
result this model is a perfect thermodynamical system
\cite{JDutta2}. It may be noted GSLT is an inherent property of
any cosmological model and it should be valid throughout the
evolution. In order to have validity of GSLT at event horizon for
self accelerating branch some interaction between dark matter and
dark DE has to be considered \cite{JDutta3}. So interaction
between dark sectors can also alleviate problems from
thermodynamical point of view.
\par Furthermore, in literature different modified DGP models have been studied. One of the simplest is the LDGP model where the source of DE is taken to be cosmological constant \cite{Lue,Lazkoz}. Some of the other modified DGP models include the following: $(i)$ QDGP model where source of DE is a quintessence field \cite{Chimento}. $(ii)$ CDGP model where source of DE is a Chaplygin Gas\cite{Lopez}. $(iii)$ SDGP model where source of DE is a scalar field\cite{Zhang}. $(iv)$ HDGP model where source of DE is a holographic dark energy \cite{Wu} etc.
\par Dynamical system study has been found to be very useful in cosmology \cite{Ellis,Coley}. The objective of dynamical system tool is to study the asymptotic behaviour of cosmological models. Stable point of the system corresponds to ultimate fate of universe. Such points are also called as late-time attractors. In other words, late time attractors give possible solution which describe our present universe irrespective of initial conditions. Recently,the interacting DE models from dynamical systems perspective have been extensively studied in literature \cite{cg,Cai,Mahata:2015nga,Biswas:2015cva,Biswas:2015zka,pathak}.
\par Scaling solutions play an important role in constructing models of DE \cite{Liddle,Scherer,Sujikawa} and are desirable
 in cosmic evolution. Here density of the scalar fields dominate at late time only and remains sub dominant at early time. In general, GR (General Relativity) based scalar field models of DE do not admit scaling attractor unless interaction is considered between dark sectors \cite{cg}. In standard cosmology scaling solution are generally unstable for phantom fields. Guo \textit{et al} have shown that it is possible to get stable scaling solutions through interacting phantom energy model\cite{Cai}.
\par Dynamical evolution of self accelerating scalar field with constant and exponential potential trapped on the DGP brane has been studied in \cite{Quiros}. This study has been extended to beyond  constant and exponential potentials by Leyva \textit{et al} \cite{Leyva}. Cosmological dynamics of quintessence and phantom field with exponential potential coupled to gravity (minimal and non-minimal) in DGP brane has been studied in \cite{Nozari}. It is noted that in all these studies, there is no late time accelerated scaling attractors. The aim of this paper is to search for late time accelerated scaling attractors in DGP braneworld.
\par  In this paper, we investigate the dynamics of scalar field (quintessence/phantom) which interacts with matter source in DGP braneworld. In GR based models, the late time accelerated scaling attractors are present only in interacting quintessence models. We shall show that it is possible to get late time accelerated scaling attractors for phantom case in DGP braneworld. Moreover, we have also investigated the classical stability of the model and found that these attractors are also classically stable.
\par The organization of the paper is as follows: In sect.\ref{DGP I:Sec2:Basic} we present the basic equations of interacting DGP braneworld model and the formation of autonomous system of differential equations. In sect.\ref{DGP I:Sec3:phase} we discuss the local and classical stability  of critical points obtained and finally, the conclusion is given in sect.\ref{sec5:conc}.
\section{Basic equation of DGP braneworld and formation of dynamical system}\label{DGP I:Sec2:Basic}
\noindent The total action of the scalar field in DGP braneworld model with matter is given by
\begin{equation} \label{DGP I:Sec2:01}
S=\frac{M_5^{3}}{2}\int d^{5}x \sqrt{-g^{(5)}} R^{(5)}+\int\left[\frac{M_{Pl}^{2}}{2} R-\theta\,\frac{1}{2}\,g^{\mu\,\nu}\nabla_\mu\phi\,\nabla_\nu\phi-V(\phi)+\mathcal{L}_{m}\right]\sqrt{-g}\,d^{4}x
\end{equation}
where $M_5$ and $M_{Pl}$ are five dimensional and four dimensional Planck mass respectively, while $g_{\mu\,\nu}^{(5)}$ and $R^{(5)}$ denote metric and Ricci scalar in the bulk respectively, the corresponding quantity in the brane are denoted by  $g_{\mu\nu}$ and $R$ respectively. Here the potential of the scalar field $\phi$ is denoted by $V(\phi)$  and $\mathcal{L}_{m}$ is a matter Lagrangian on the brane. Further we note that for $\theta=1$, we get an ordinary (quintessence) scalar field and $\theta=-1$ corresponds to a phantom field. \\
\par Observations support spatially flat \cite{Miller} Friedmann Robertson Walker (FRW) spacetime, given by the line element
\begin{equation}\label{DGP I:Sec2:02}
 ds^{2}= - dt^{2}+a^{2} (t) (dx^{2}+dy^{2}+dz^{2})
\end{equation}
where $a(t)$ is a scale factor.\\
\par If we vary (\ref{DGP I:Sec2:01}) with respect to metric tensor components, then we get modified Friedmann equation of DGP model in the above spacetime \cite{dgp1,dgp2} as\\
\begin{equation}\label{DGP I:Sec2:03}
H^{2}-\epsilon\frac{H}{r_{c}}=\frac{\rho}{3}
\end{equation}\\
where $\epsilon=\pm 1$, $H=\frac{\dot{a}}{a}$ is the Hubble parameter and $r_c=\frac{M_{Pl}^2}{2 M_5^3}$ is known as the cross-over scale which differentiates the brane dynamics of universe from the usual $4$D universe.\\
\par The usual $4$D Friedmann equation is obtained when $H^{-1}<<r_{c}$, but  $H^{-1}>>r_{c}$ implies the 5-dimensional effect of gravity.  Further $\epsilon=1$ corresponds to DGP(+) model which is self-accelerating, while for $\epsilon=-1$ we have DGP(-) model which requires DE on the brane to accelerate. In this paper, we study DGP(-) model where the scalar field (quintessence/phantom) is taken as source of DE.\\
~ Eq. (\ref{DGP I:Sec2:03}) can also be written as
\begin{equation}\label{DGP I:Sec2:04}
H^{2}+\frac{H}{r_{c}}=\frac{\rho_m+\rho_\phi}{3}
\end{equation}\\
where total energy is taken as $\rho_{\phi}+\rho_m$. Here $\rho_{\phi}$ is energy density of scalar field and $\rho_m$ is the energy density of the matter (Baryonic+DM). The matter is taken as a perfect fluid with barotropic equation of state $p_m=(\gamma-1)\,\rho_m$ where a constant $\gamma$ is known as barotropic index of perfect fluid $(0\leq\gamma\leq2)$. Since DM is the dominant source of matter, therefore for brevity we denote matter source by DM.\\
\par Eq. (\ref{DGP I:Sec2:04}) can be written as
\begin{equation}\label{DGP I:Sec2:04a}
H^2=\frac{1}{3}(\rho_{m}+\rho_{\rm eff})
\end{equation}
where
\begin{equation}\label{DGP I:Sec2:04b}
\rho_{\rm eff}=\rho_{\phi}-\frac{3H}{r_c}
\end{equation}
~ The energy density and pressure of a scalar field are respectively given by
\begin{equation}\label{DGP I:Sec2:05}
 \rho_{\phi}=\theta \, \frac{1}{2}\dot{\phi}^{2}+V(\phi)
\end{equation}
\begin{equation}\label{DGP I:Sec2:06}
p_{\phi}=\theta\, \frac{1}{2}\dot{\phi}^{2}-V(\phi)
\end{equation}
~ The energy conservation equations for $\rho_m$, $\rho_{\phi}$  are respectively given by
 \begin{equation}\label{DGP I:Sec2:07}
 \dot{\rho}_{m}+3\,H\,\gamma\,\rho_{m}=-Q
\end{equation}
\begin{equation}\label{DGP I:Sec2:08}
\dot{\rho}_{\phi}+3\,H\,(\rho_{\phi}+p_{\phi})=Q
\end{equation}\\
 where $Q$ is the strength of interaction between DE and DM. The sign of $Q$ determines the direction of energy transfer. For $Q>0$, energy is transferred from DM to DE and for $Q<0$ energy is transferred from DE to DM.  For $Q=0$, $\theta=1$, the study reduce to the case of non-interaction which had been studied in literature \cite{Quiros,Leyva,Nozari}.

 Furthermore, we have conservation equation for effective energy density given by
\begin{equation}\label{DGP I:Sec2:08a}
\dot{\rho}_{\rm eff}+3H(1+\omega_{\rm eff})\rho_{\rm eff}=Q
\end{equation}
where $\omega_{\rm eff}=\frac{p_{\rm eff}}{\rho_{\rm eff}}$.\\

From the eqs. (\ref{DGP I:Sec2:04}), (\ref{DGP I:Sec2:07}) and (\ref{DGP I:Sec2:08}) we obtain
\begin{equation}\label{DGP I:Sec2:11a}
\frac{\dot{H}}{H^2}=-\frac{3}{2}\left[\frac{(1+\omega_{\phi})\Omega_{\phi}+\gamma\,\Omega_m}{1+\sqrt{\Omega_{r_c}}}\right]
\end{equation}
~ From eqs. (\ref{DGP I:Sec2:04b}), (\ref{DGP I:Sec2:08}), (\ref{DGP I:Sec2:08a}) and (\ref{DGP I:Sec2:11a}) we obtain
\begin{equation}\label{DGP I:Sec02:08b}
1+\omega_{\rm eff}=\frac{\sqrt{\Omega_{r_c}}\left[(1+\omega_{\phi})\Omega_{\phi}-\gamma\Omega_m\right]}{\Omega_{\rm eff}}
\end{equation}
where
\begin{equation}
\Omega_{\rm eff}=\frac{\rho_{\rm eff}}{3\,H^2}
\end{equation}
From eq. (\ref{DGP I:Sec2:04}),
\begin{equation}\label{DGP I:Sec2:09}
1=\Omega_{m}+\Omega_{\phi}-2\sqrt{\Omega_{r_{c}}}
\end{equation}
where $\Omega_{m}=\frac{\rho_{m}}{3H^{2}}$ is dimensionless matter energy density parameter, $\Omega_{\phi}=\frac{\rho_{\phi}}{3H^{2}}$ is dimensionless dark energy density parameter and $\Omega_{r_{c}}=\frac{1}{4r_{c}^2H^2}$ is dimensionless parameter which determines the DGP character.\\\linebreak
~ Using eqs. (\ref{DGP I:Sec2:05}) and  (\ref{DGP I:Sec2:06}) in eq.(\ref{DGP I:Sec2:08}), the equation of motion of scalar field is obtained as
\begin{equation}\label{DGP I:Sec2:10}
\ddot{\phi}+3H\dot{\phi}+\theta\,\frac{dV}{d\phi}=\frac{\theta\,Q}{\dot{\phi}}
\end{equation}
~ We now introduce the following dimensionless variables
\begin{equation} \label{DGP I:Sec2:11}
x=\frac{\dot{\phi}}{\sqrt{6}H}, \qquad y=\frac{\sqrt{V}}{\sqrt{3}H},\qquad z=\frac{1}{\sqrt{2r_{c}H}}\qquad s=-\frac{1}{V}\frac{dV}{d\phi}
\end{equation}
\par Eq. (\ref{DGP I:Sec2:11}) implies that at $z=0$ corresponds to $r_c\rightarrow\infty$, in which brane effect will vanish and it reduces to standard $4$ dimensional behaviour.\\
\par The relevant cosmological parameters in terms of dimensionless variables (\ref{DGP I:Sec2:11}) $viz.,$ DM energy density parameter, DE density parameter, equation of state parameter for scalar field and deceleration parameter are respectively given by
\begin{eqnarray}
\Omega_{m}&=&1-\theta\,x^{2}-y^{2}+2z^{2} \label{DGP I:Sec2:12}\\
\Omega_{\phi}&=&\theta\, x^{2}+y^{2} \label{DGP I:Sec2:13}\\
\omega_{\phi}&=&\frac{\theta \, x^{2}-y^{2}}{\theta\, x^{2}+y^{2}} \label{DGP I:Sec2:14}\\
q&=&-1+\frac{3}{2(z^{2}+1)}\left(2x^{2}+\gamma(1-\theta\,x^{2}-y^{2}+2z^{2})\right)\label{DGP I:Sec2:15}
\end{eqnarray}
\par Since $0\leq\Omega_{m}\leq1$, so from eq. (\ref{DGP I:Sec2:12}), we have $\theta\,x^{2}+y^{2}\leq 2z^{2}+1$. It may be noted that eqns.(\ref{DGP I:Sec2:12}-\ref{DGP I:Sec2:15}) coincide with those of ref \cite{Nozari} for $\theta=1$ and $Q=0$. \\
\par We now estimate the initial conditions for numerical solutions in such a way that it matches with present observational data ($\Omega_{r_c}=0.12,\Omega_m=0.27$)\cite{Liang} and present observed value of deceleration parameter $q_0=-0.61$ \cite{Chao}.  Using eqns. (\ref{DGP I:Sec2:12}) and (\ref{DGP I:Sec2:15}), yield the following lower bound set for the present work.
\begin{equation}\label{DGP I:Sec2:16}
 x_0=\pm0.20,\,\, y_0=\pm1.17, \,\, z_0=\pm 0.59 \qquad
\end{equation}
~ Here, $x_0$, $y_0$ and $z_0$ are present values of $x$, $y$ and $z$ respectively  (\textit{i.e.,}$\,N=\ln\,a=0$).
\section{Phase space analysis}\label{DGP I:Sec3:phase}
\label{sec3}
\noindent This section deals with  local and classical stability analysis of critical points of a corresponding autonomous system and their cosmological implications. Before going to the discussion of stability of critical points, we review briefly some methods that will be used in this paper.\\
Let $\mathbf{x'}=\mathbf{f}(\mathbf{x})$ denotes a non-linear autonomous system and $\mathbf{x_*}$ be a critical point \textit{i.e.,} $\mathbf{f}\,(\mathbf{x_*})=0$ where $\mathbf{f}:\mathbb{R}^n\rightarrow\mathbb{R}^n$. The linearised form of a given non-linear system near a critical point can be written as  $\mathbf{x'}=A\mathbf{x}$ where $A=\textbf{Df}(\mathbf{x_*})$ is the Jacobian matrix of $\mathbf{f}$ at $\mathbf{x_*}$ and $\textbf{Df}(\mathbf{x_*})=\left(\frac{\partial\,f_i}{\partial\,x_j}\right)$, $i,j=1,2,3...,n$.
A critical point is called a \textit{hyperbolic point} if all the eigenvalues of its corresponding Jacobian matrix contains a \textit{non-zero} real components.
 In this case one can apply linear stability analysis to check the stability of a point \cite{Wiggins}.  Otherwise it is a \textit{non-hyperbolic point} where linear stability theory
  cannot give any valid stability decisions. The perturbation plot is very popular now a days to determine the stability of non hyperbolic critical points \cite{Strogatz}.
   Other known mathematical tools like Centre Manifold Theory, Lyapunov functions \cite{Wiggins,Perko} can also be used to determine the stability of such critical point.
    For a set of non-isolated critical points, if its Jacobian matrix contains only one eigenvalue with zero real part and the rest are all non-zero and the
     eigenvector associated with a zero eigenvalue is tangent to the set of critical points then the set is said to be \textit{normally hyperbolic set} \cite{Aulbach}.
     The stability of this set is determined by the signs of the remaining non-zero eigenvalues.  In this present work for non-hyperbolic points we use numerical methods
     of perturbed solutions around a critical point and normally hyperbolic set property. These methods have been extensively used  recently in studying cosmological scenarios\cite{Nandan,Jibitesh}.\\
In what follows we choose following two types of specific interactions
\begin{center}
(A)\quad $Q=\sqrt{\frac{2}{3}}\,\alpha\,\rho_m \,\dot{\phi}$\qquad \qquad
(B)\quad $Q=\beta\,\dot{\rho}_{\phi}$
\end{center}
While interaction A was introduced in \cite{Wetterich,Amendola} and a dynamical study in the context of GR both quintessence and phantom was studied in \cite{cg,Billiyard}, interaction B was studied recently for quintessence in GR context \cite{pathak} where coincidence problem is alleviated in comparison to the uncoupled model \cite{Liddle}. These interactions belong to the class of local interactions where both depend directly  on the energy density unlike the class of non-local/global interactions which depends on the Hubble parameter and energy density \cite{cg}. \\\linebreak
In order to determine the stability of critical points of an autonomous system, we need to specify form of potentials. The consideration of a specific potential is generally done by ad hoc mechanism. In literature various candidates have been proposed such as inverse power law, exponential, hyperbolic and many more (for review see \cite{cope, Amen}). Depending on the choice of $\Gamma\equiv\frac{V\,\frac{d^2V}{d\phi^2}}{\left(\frac{dV}{d\phi}\right)^2}$, the potential can be broadly categorised as\cite{NandanRoy}:\\
\textbf{Category I}: Non-exponential form, $\Gamma\neq1$.\\
In order to study the nature of non-hyperbolic points, we consider one concrete potential $V(\phi)=\frac{M^{4+n}}{\phi^n}$, where $M$ and $n$ are constants for which
 $\Gamma=1+\frac{1}{n}$. This potential can lead to tracking behaviour \cite{Steinhardt}. Tracker field is very important from coincidence problem point of view as scalar field tracks the background matter energy density throughout the history of the universe and eventually overtakes the matter density to produce late time acceleration.\\
\textbf{Category II}: Exponential form, $\Gamma=1$\\
Exponential potential of scalar field models arise naturally from fundamental theories such as String theory/M-Theory\cite{Billy_thesis}. Exponential potential
 produce scaling solutions which are desirable from coincidence problem point of view. Exponential potential plays a crucial cosmological role for driving cosmological
  inflation period \cite{Lucchin,Wett,Wands}. In case of exponential potential, we consider $V=V_0\,\exp\,(-\lambda\,\phi)$. Here we assume $\lambda>0$ as $\lambda<0$ can
   be associated with the change of $\phi\rightarrow\,-\,\phi$. The advantage of this potential is that the phase space reduce to three dimension, from which behaviour of a system can
    be easily studied.
\par Local stability of a point is related to small perturbations on values of $x, y, z$ and $s$ near a point. Classical stability of a model is related to the fluctuation in dark energy pressure $\delta p_{\phi}$. In cosmological perturbation theory, the important quantity which plays a key factor for stability of classical fluctuation is the adiabatic speed of sound $C_s^2$ defined by $C_s^2=\frac{\partial p/ \partial N}{\partial \rho/\partial N}$. The model is said to be classically stable if $C_s^2\geq 0$ at local critical points\cite{Piazza,Mahata}.  It is important to note that  local stability does not imply the classical stability. From cosmological point of view, those points which are locally as well as classically stable are of interest.
\subsection{\textbf{Interaction A: $Q=\sqrt{\frac{2}{3}}\,\alpha\,\rho_m\, \dot{\phi}$}}\label{intI}
\noindent In this case using dimensionless variables (\ref{DGP I:Sec2:11}) the evolution equations can be converted to the following autonomous system

\begin{eqnarray}
x'&=&-3\,x+\frac{\theta}{2}\,s\,\sqrt {6}\,{y}^{2}+\frac{3}{2}\,{\frac {x \left( 2\,\theta\,{x}^{2}+\gamma\, \left( 2\,{z}^{2}-\theta\,{x}^{2}-{y}^{2}+1 \right)  \right) }{{z}^{2}+1}}+\theta\, \alpha\, \left( 2\,{z}^{2}-\theta\,{x}^{2}-{y}^{2}+1 \right) \label{DGP I:Sec3:17} \\
y' &=& -\frac{1}{2}\,s\sqrt {6}\,x\,y+\frac{3}{2}\,{\frac {y \left(2\,\theta\,{x}^{2}+\gamma\, \left( 2\,{z}^{2}-\theta\,{x}^{2}-{y}^{2}+1 \right)  \right) }{{z}^{2}+1}}\label{DGP I:Sec3:18} \\
z'&=& \frac{3}{4}\,{\frac {z \left(2\,\theta\,{x}^{2}+\gamma\, \left( 2\,{z}^{2}-\theta\,{x}^{2}-{y}^{2}+1 \right)  \right) }{{z}^{2}+1}}\label{DGP I:Sec3:19}\\
s'&=& -\sqrt{6}\,x\,{s}^{2}\left(\Gamma-1\right) \label{DGP I:Sec3:20}
\end{eqnarray}
where prime denotes derivative with respect to $N=\ln a$. It is noted from eqn.(\ref{DGP I:Sec3:17}) that when $\Omega_m=0$, the last term containing $\alpha$ vanishes\textit{ i.e.,} point where $\Omega_{m}=0$ is independent of the interaction $Q$ for its existence and hence it exists in case of uncoupled model also. It can be seen that the above system  is invariant under the change of sign $\,z\rightarrow\,-z$ and $\,y\rightarrow\,-y$. So, we restrict our analysis to the positive values of $y$ and $z$ only.\\\linebreak
The adiabatic speed of sound $C_s^2$ is given by
\begin{equation}
C_s^2=1+\frac{2sy^2}{\left(\sqrt{\frac{2}{3}}\alpha(1-\theta\,x^2-y^2+2z^2)-\sqrt{6}\,\theta\,x\right)} \label{DGP I:Sec2:21}
\end{equation}
In what follows we study the phase space analysis of above two categories of potentials separately.
\subsubsection{\textbf{Category I:}\textit{Non-exponential form of potential} ($\Gamma\neq1$)}  \label{sub3,I}
\noindent In this category, eqns. (\ref{DGP I:Sec3:17})-(\ref{DGP I:Sec3:20}) form a closed system of equations. The critical points along with corresponding cosmological parameters are given in table \ref{Tab I} and the eigenvalues of their corresponding Jacobian matrix 
are given in table \ref{Tab II}.
\begin{center}
\begin{table}[h!]
\caption{Critical points and corresponding cosmological parameters}.
\begin{center}
\begin{tabular}{c c c c c c c c c}
\hline\hline
Point  &  $~~~x~~~$   &$~~~y~~~$    &$~~~z~~~$  &$~~~s~~~$    & ~~~Existence    ~~~&$~~~\Omega_\phi~~~$    &$~~~\omega_\phi~~~$   &$~~~q~~~$ \\ \hline\\
$A_{1}$&  $\frac{1}{\sqrt{\theta}}$   &$0$    &$0$   &$0$    &$\theta>0$        & $1$             & $1$            &$2$ \\ [1.5ex]
$A_{2}$&  $-\frac{1}{\sqrt{\theta}}$  &$0$    &$0$   &$0$    &$\theta>0$        & $1$             & $1$            &$2$ \\ [1.5ex]
 & & & & & $\gamma \neq 2$ and & & & \\ [-1ex]
 \raisebox{1.5ex}  {$A_3$} & \raisebox{1.5ex}{$ \frac{2 \alpha\,\theta}{3(2-\gamma)}$}& \raisebox{1.5ex}{$0$}&\raisebox{1.5ex}{$0$}&\raisebox{1.5ex}{$0$}& $\theta \alpha^{2}<\frac{9(2-\gamma)^{2}}{4}$ & \raisebox{1.5ex}{ $\frac{4 \theta\,\alpha^2}{9(2-\gamma)^2}$} &\raisebox{1.5ex}{ $1$} &\raisebox{1.5ex}{$\frac{2 \alpha^2\theta}{3 (2-\gamma)}+\frac{3}{2}\gamma  -1$} \\ [1.5ex]
$A_4$ & $0$&$\sqrt{2\,z^{2}+1}$&$z$&$0$&Always  & $2\,z^{2}+1$ & $-1$&$-1$\\[1.5ex]\hline\hline
\end{tabular}\label{Tab I}
\end{center}
\end{table}
\end{center}
\begin{center}
\begin{table}[h!]
\caption{Eigenvalues of critical points in table \ref{Tab I} }
\begin{center}
\begin{tabular}{ccccccc}
\hline \hline
Point~~~& ~~~$E_1$~~~ & ~~~$E_2$~~~ & ~~~$E_3$~~~ & $~~~E_4~~~$&$~~~C_s^2~~~$ \\ \hline \\
$A_{1}$ & $0$ & $\frac{3}{2}$ & $3$ & $3(2-\gamma)-2\,\alpha\sqrt{\theta}$& $1$
\\ [1.5ex]
$A_{2}$ & $0$ & $\frac{3}{2}$ & $3$ & $3(2-\gamma)+ 2\,\alpha\sqrt{\theta}$&$1$
 \\ [1.5ex]
$A_3$ & 0 & $ \frac{4\,\alpha^{2}\theta+9\,\gamma\,(2-\gamma)}{12\,(2-\gamma)} $ &$\frac{4\,\alpha^{2}\theta-9\,(2-\gamma)^{2}}{6\,(2-\gamma)} $ & $\frac{4\,\alpha^{2}\theta+9\,\gamma\,(2-\gamma)}{6\,(2-\gamma)} $&$1$
\\ [1.5ex]
$A_4$ & $0$ & $0$ & $-3$ & $-3\,\gamma$&stable (limiting)
\\ [1.5ex] \hline\hline
\end{tabular}\label{Tab II}
\end{center}
\end{table}
\end{center}
We now discuss the stability of critical points given in table \ref{Tab I} separately for quintessence and phantom field.\\\linebreak
(i) Quintessence field ($\theta=1$):\\
Points $A_1$ and $A_2$ correspond to the un-accelerated, dominated by kinetic part of quintessence field (\,$\Omega_{\phi}=1,\Omega_m=0, q=2$). $A_1$ is an unstable node if $3(2-\gamma)>2\alpha$ and $A_2$ is an unstable node if $3(2-\gamma)>-2\alpha$, else they behave as saddle points. Point $A_3$ behaves as a saddle fixed point (since $E_2$ and $E_4$ are positive but $E_3$  is negative in the existence region of this point). Points on a set of critical points $A_4$ correspond to an accelerated solution ($q=-1$) and since it has two zero eigenvalues and two negative eigenvalues, so linear stability theory is not enough and further investigation is required to decide the stability of this set.
\begin{figure}
\centering
\subfigure[]{%
\includegraphics[width=8cm,height=6cm]{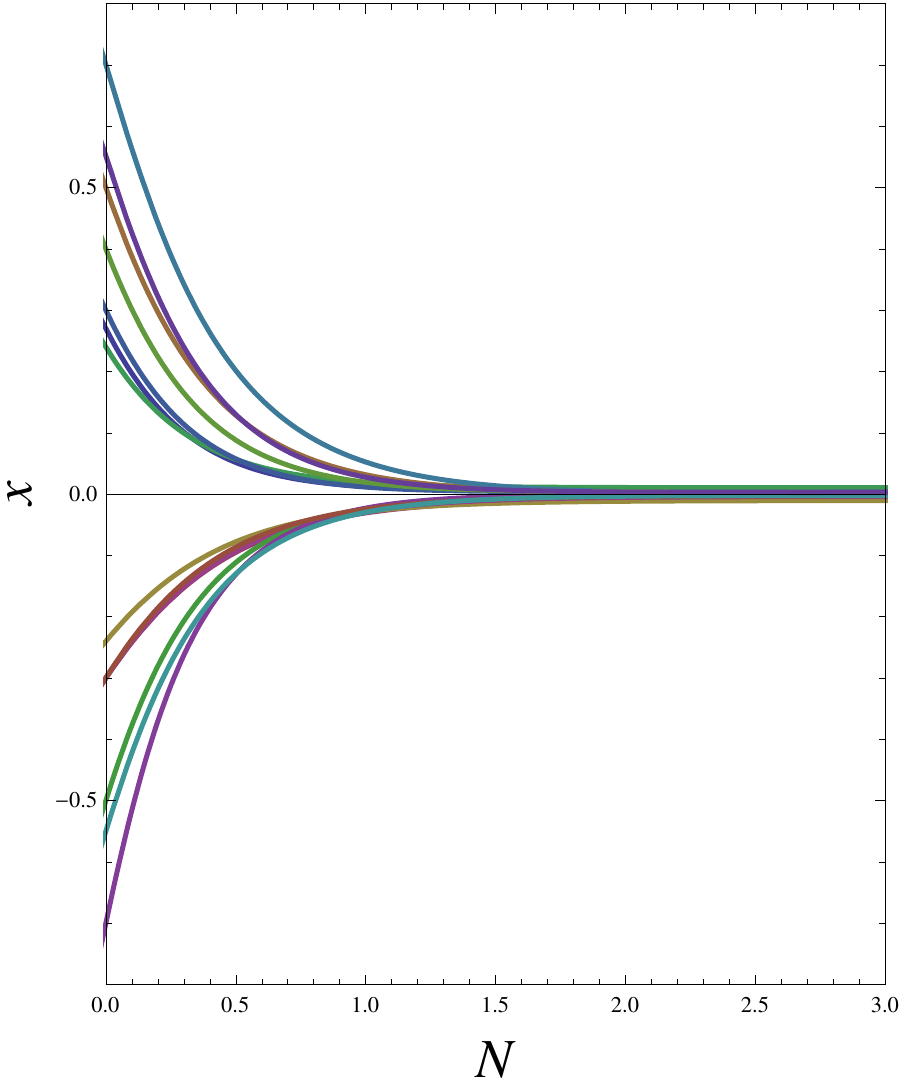}\label{fig1}}
\qquad
\subfigure[]{%
\includegraphics[width=8cm,height=6cm]{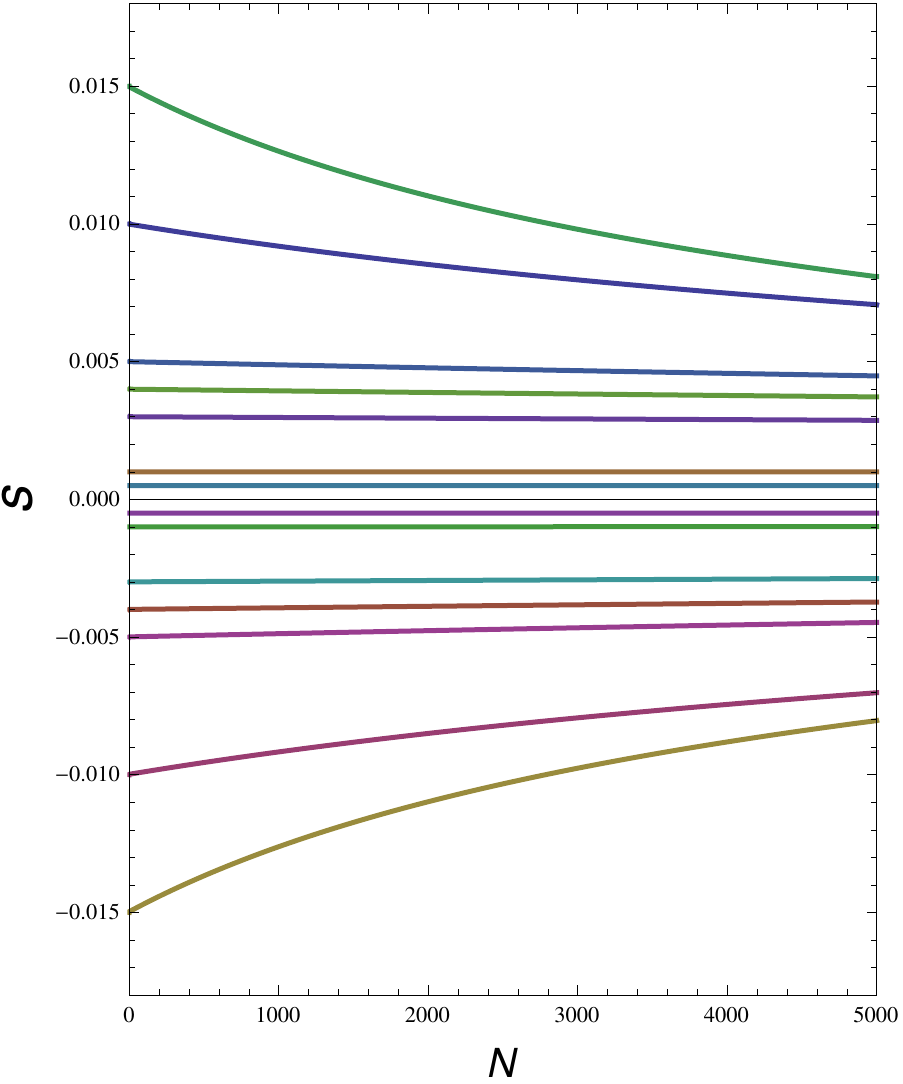}\label{fig4}}\\
\qquad
\subfigure[]{%
\includegraphics[width=8cm,height=6cm]{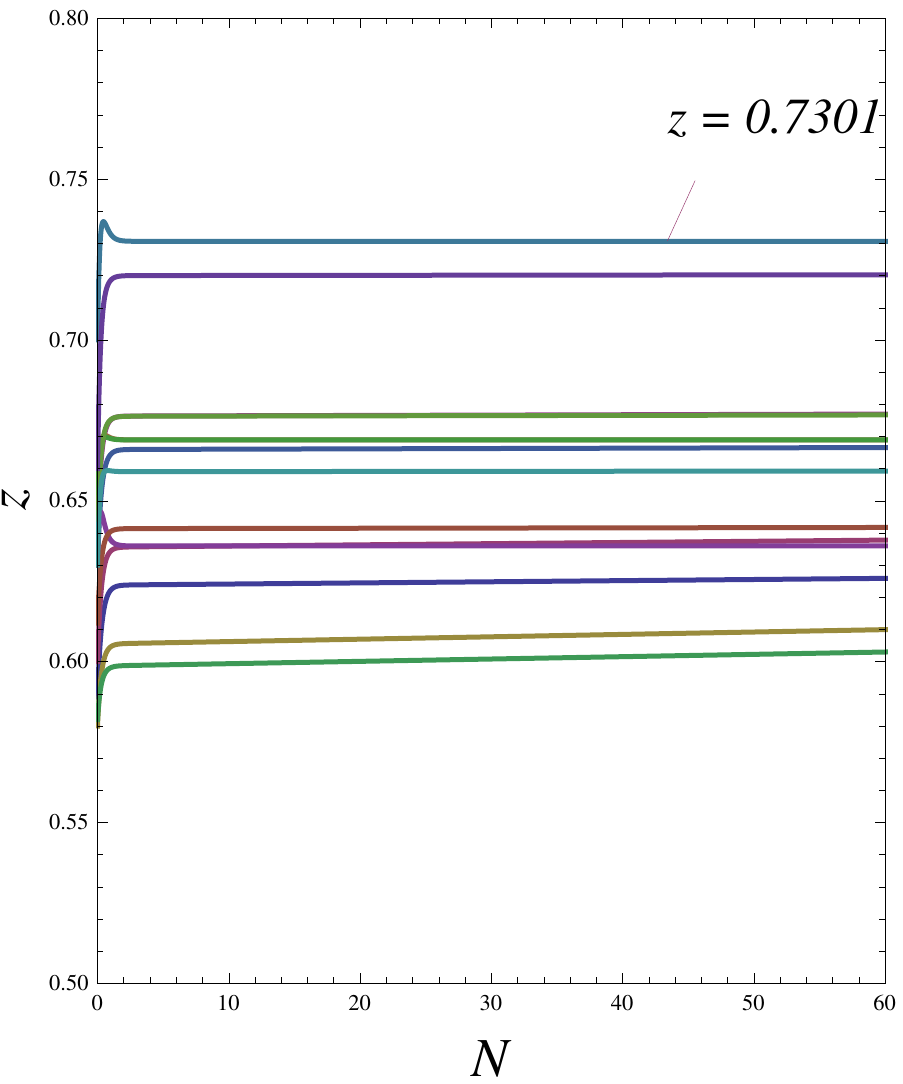}\label{fig3}}
\qquad
\subfigure[]{%
\includegraphics[width=8cm,height=6cm]{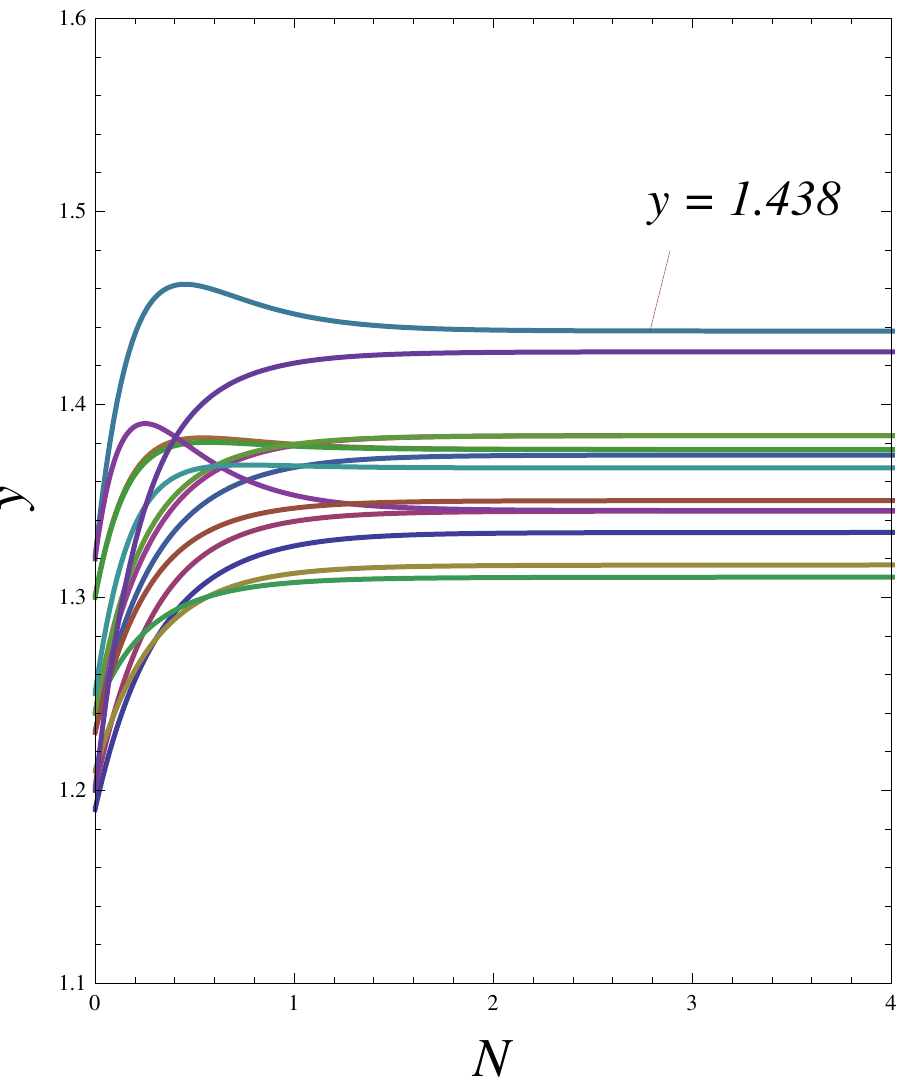}\label{fig2}}

\caption{(a). Projection of perturbation plot of $x$ versus $N$.  (b). Projection of perturbation plot of $s$ versus $N$. (c).
Projection of perturbation plot of $z$ versus $N$. (d). Projection of perturbation plot of $y$ versus $N$ for $\theta=1$.}
\end{figure}

To check the stability of this set of non isolated critical points $A_4$, we numerically perturb the solutions around a critical point of the set.

We plot the perturbation plots projected on the $x,y,z$ and $s$
axes separately. From figs.\ref{fig1} and \ref{fig4}, it is
evident that trajectories of perturbed solutions approach $x=0$
and $s=0$ respectively as $ N\rightarrow\infty$ . It seems that
trajectories in fig.\ref{fig4} are parallel to a horizontal axis
but they do converge to $s=0$, but converge slowly. Indeed we have
checked that trajectories actually converge to $s=0$ as
$N\rightarrow\infty$. Furthermore, we note from fig.\ref{fig3}
that any perturbation of the system near $z$ makes it constant at
the perturbed value and it shows that $z$ is arbitrary. We can
also see from fig.\ref{fig2}, that for each value of $z$, where
the trajectories approach as $ N\rightarrow\infty$, the
corresponding trajectories of $y$ also approach the value
$\sqrt{2z^2+1}$ as $ N\rightarrow\infty$. One such trajectory is
shown in fig.\ref{fig3} where $z=0.7301$ and
$y=\sqrt{2z^2+1}=1.438$ in fig.\ref{fig2}. From these behaviours
of the system near $A_4$, we can conclude that $A_4$ is a late
time attractor. It is interesting to see the effect of brane in
solution $A_4$. This indeed shows role of brane in explaining late
time acceleration.
\par In the phase space of the autonomous system any heteroclinic orbit starts from an unstable critical point (past time attractor) and evolve to a stable critical point (late time attractor) via saddle points.
\par So, a viable cosmological model must have a past time attractor, saddle points and late time attractors to represent early universe, radiation or matter dominated eras and late time acceleration respectively.
\par In this case, universe evolves from one of these unstable points $A_1$ or $A_2$ and approaches toward the saddle point $A_3$ and finally settles down towards the attractor set $A_4$. Furthermore, the attractor set $A_4$ is also classically stable and very interesting from cosmological point of view.\\

\begin{figure}
\centering
\subfigure[]{%
\includegraphics[width=7cm,height=5cm]{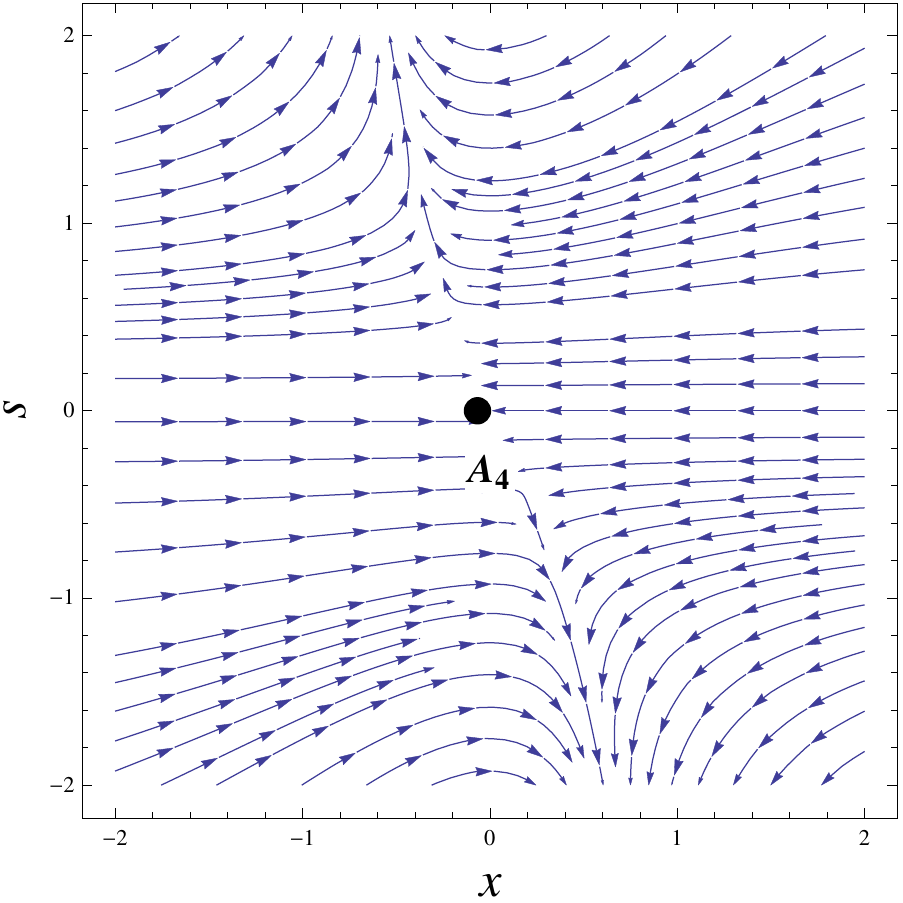}\label{sppb}}
\qquad
\subfigure[]{%
\includegraphics[width=7cm,height=5cm]{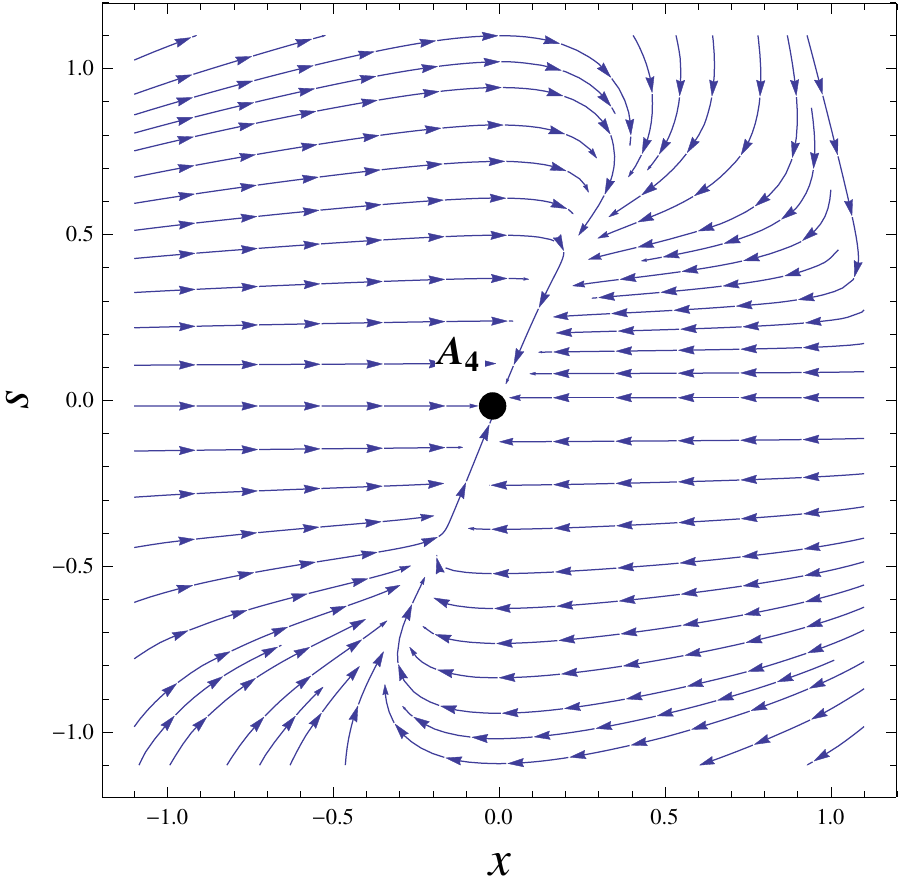}\label{spqb}}
\caption{(a). Projection of the system (\ref{DGP I:Sec3:17})-(\ref{DGP I:Sec3:20}) on $x-s$ plane for $\theta=-1$. (b). Projection of the system  (\ref{DGP I:Sec3:17})-(\ref{DGP I:Sec3:20}) on $x-s$ plane for $\theta=1$. Here $\alpha=-0.7$, $\gamma=1$.}
\end{figure}

\noindent (ii) Phantom field ($\theta=-1$):\\
Critical points  $A_1,\, A_2$ do not exist for this case of phantom field. Point $A_3$ is physically meaningless, since $\Omega_{\phi}$  is negative. The set of critical points $A_4$ corresponds to an accelerated solution. Fig.\ref{sppb} shows the 2D projection of the system on the $x-s$ plane. We observe that trajectories which initially approach a point $(0,0)$ in $x-s$ plane, ultimately moves away from it. This implies that a set of critical points $A_4$ is an unstable set unlike the case of a quintessence field which is stable (see figs.(\ref{sppb} and \ref{spqb}). Thus, non exponential potential do not give any interesting cosmological scenarios for this case.

\subsubsection{\textbf{Category II}: \textit{Exponential form of potential} ($\Gamma=1$)} \label{sub3,2}
\begin{figure}[h!]
\centering
\subfigure[]{%
\includegraphics[width=7cm,height=5cm]{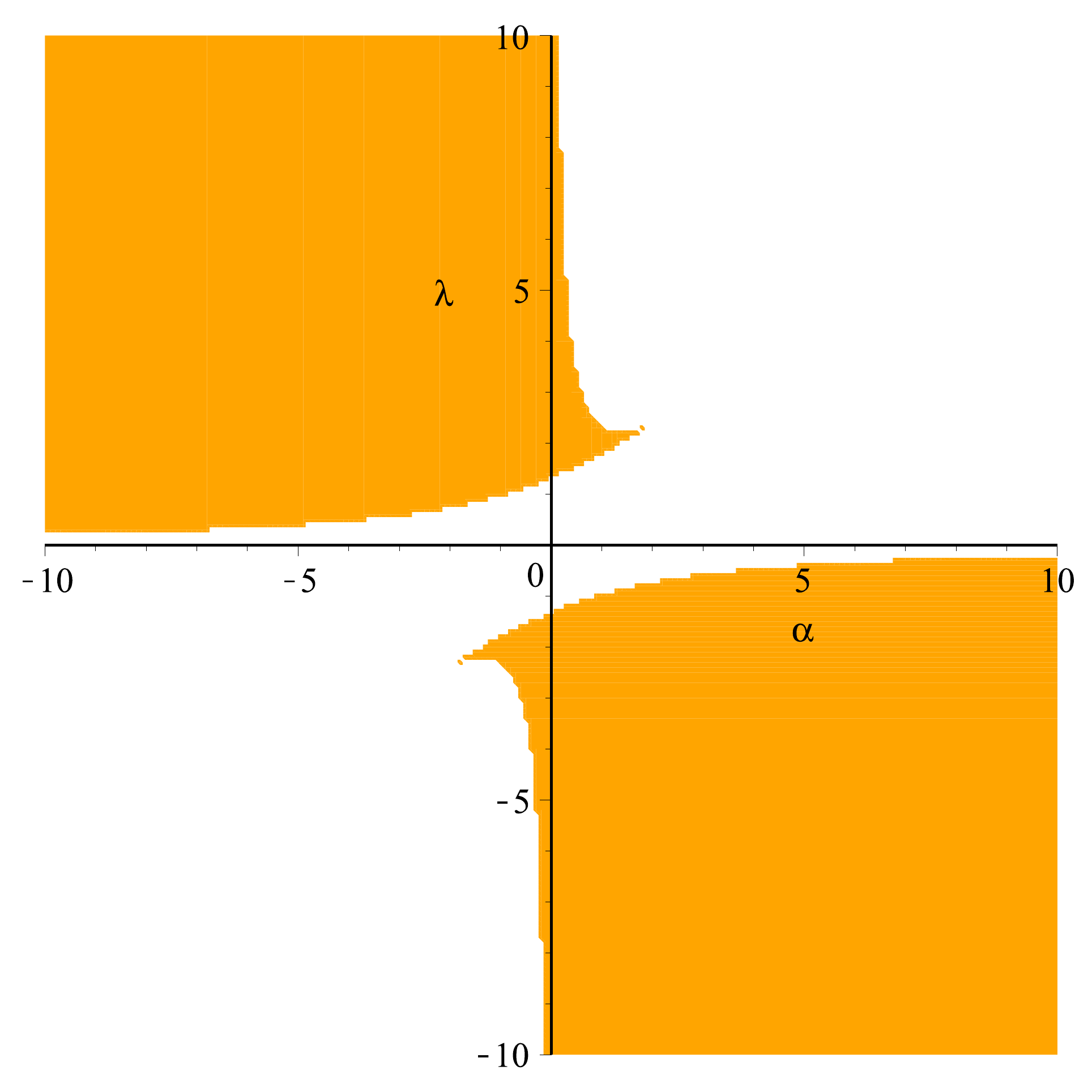}\label{stb5}}
\qquad
\subfigure[]{%
\includegraphics[width=7cm,height=5cm]{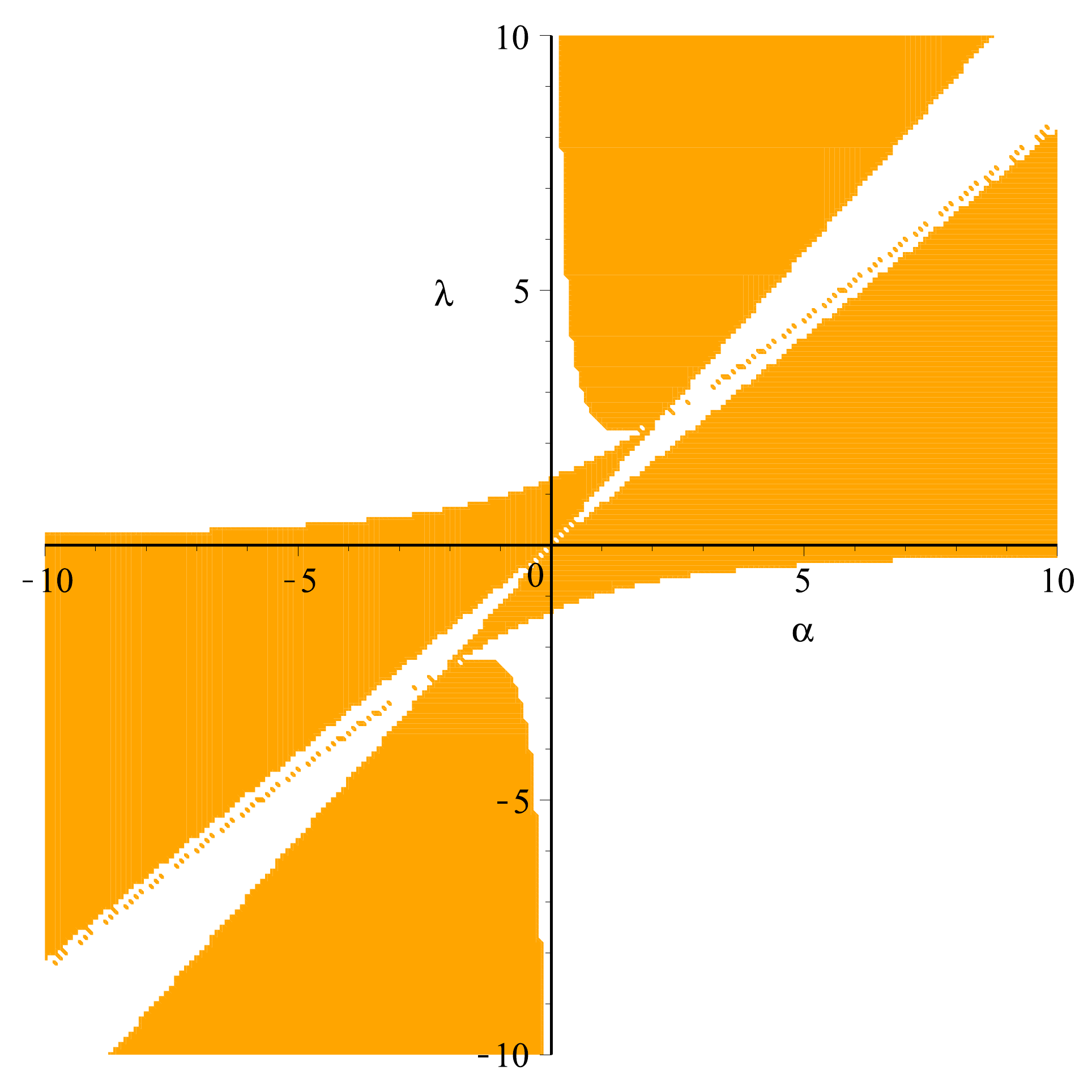}\label{pe3b5}}
\caption{(a). Stability region of $B_5$. (b). Region of positivity
of eigenvalue $E_3$ with $\theta=1\,,\gamma=1$ for $B_5$.}
\end{figure}

\noindent In this category, $s$ is constant  and $V=V_0\, \rm {exp}(-\lambda\phi)$, so eqs.(\ref{DGP I:Sec3:17})-(\ref{DGP I:Sec3:19}) form a closed system of equations. Critical points and their cosmological parameters are listed in table \ref{Tab III} and the eigenvalues of their corresponding Jacobian matrix are given in table \ref{Tab IV}. In what follows we discuss the stability of critical points for $\theta = 1$ and $\theta=-1$ separately under this potential.\\\linebreak
(i) Quintessence field ($\theta=1$):\\
Point $B_1$ is an unstable node if $\alpha<\frac{3(2-\gamma)}{2}$ and $\lambda<\sqrt{6}$, otherwise it is saddle. $B_2$ is an unstable node if $\alpha<\frac{3(2-\gamma)}{2}$ and $\lambda>-\sqrt{6}$, otherwise it is saddle. Both $B_1, B_2$ correspond to un-accelerated, quintessence kinetic energy dominated solutions ($q=2,\Omega_{\phi}=1$). Point $B_3$ corresponds to a scaling solution for $\alpha\neq 0$ and it is a saddle point (since eigenvalue $E_2$ is negative, whereas   $E_1$ is always positive in its region of existence). Point $B_4$ corresponds to scalar field dominated point, which can be accelerated if $\lambda^2<2$. It is a saddle point since $E_1$ is positive and $E_2$ is negative in the region of existence. From figs.\ref{stb5}  and \ref{pe3b5} it can be seen that the region of existence of $B_5$ and region of positivity of its eigenvalue $E_3$ are disjoint. This numerically confirms that $E_3$ is negative, but $E_1$ is always positive. Thus, scaling point $B_5$ is a saddle point . The unstability of $B_5$ is in contrary with the result in standard GR found in references \cite{cg,Billiyard} where this point corresponds to a scaling late time attractor. The set of critical points $B_6$  demands $\lambda=0$ for its existence, which means that $V(\phi)$ is constant. It is a normally hyperbolic set.  Since the remaining non-zero eigenvalues are all negative, so the set of critical points $B_6$ is a late time attractor.
\begin{center}
\begin{table}[h!]
\caption{ Critical points and their cosmological parameters of system (We have defined:\,$ b=\lambda-\frac{\sqrt{6}}{3}\alpha$) }
\begin{center}
\begin{tabular}{cccccccc}
\hline\hline
~~Point~~&$~~x~~$&$~~~y~~~$&$~z~$&~~~Existence~~~&$~~~~~~~~\Omega_\phi~~~~~~~~$&$~~~\omega_\phi~~~$&$~q$ \\ \hline\\
$B_{1}$&$\frac{1}{\sqrt{\theta}}$&$0$&$0$&$\theta>0$  & $1$ & $1$&$2$ \\ [2ex]
$B_{2}$&$-\frac{1}{\sqrt{\theta}}$&$0$&$0$&$\theta>0$  & $1$ & $1$&$2$ \\ [2ex]
& & & & $ \gamma \neq 2 $ & & &  \\ [-1ex]
\raisebox{1.5ex}{$B_3$} & \raisebox {1.5ex} {$ \frac{2\theta \alpha}{3(2-\gamma)}$} & \raisebox {1.5ex} {$0$} &\raisebox {1.5ex}{$ 0 $}& $\theta \alpha^{2}<\frac{9(2-\gamma)^{2}}{4}$ & \raisebox {1.5ex}{$\frac{4\theta \alpha^2}{9(2-\gamma)^2}$} & \raisebox {1.5ex}{$1$} & \raisebox {1.5ex}{$\frac{2\theta \alpha^2}{3 (2-\gamma)}+\frac{3}{2}\gamma  -1$} \\ [2ex]
$B_4$&$ \frac{\theta\lambda}{\sqrt{6}}$&$\sqrt{1-\frac{\theta\lambda^{2}}{6}}$&$0$&$\theta\lambda^2<6$&$1$&$\frac{\theta\lambda^2}{3}$&$\frac{\theta\lambda^2}{2}-1$\\ [2ex]
& & & & $9\theta\gamma(2-\gamma)>2\sqrt{6}\alpha\,b$ & & & \\[-1ex]
\raisebox{1.5ex}{$B_5$} & \raisebox{1.5ex}{$\frac{\gamma\sqrt{6}}{2\,b}$} & \raisebox{1.5ex}{$\frac{\sqrt{9\theta\gamma(2-\gamma)-2\sqrt{6}\alpha\, b}}{\sqrt{6}\,b}$} & \raisebox{1.5ex}{$0$} & $9\gamma^2+9\theta\gamma(2-\gamma)<2b(\sqrt{6}\alpha+3b)$ & \raisebox{1.5ex}{$\frac{9\gamma(\gamma+\theta(2-\gamma))-2\sqrt{6}\alpha b}{6 b^2}$} & \raisebox{1.5ex}{$\frac {9\theta{\gamma}^{2}+\sqrt {6}\alpha b-9\theta\gamma}{-\sqrt {6}\alpha b+9\theta\,\gamma}$} & \raisebox{1.5ex}{$\frac{\sqrt{6}\gamma\alpha+3\gamma b-2b}{2b}$} \\ [2ex]
$B_6$ & $0$&$\sqrt{2\,z^{2}+1}$&$z$&Always  & $2\,z^{2}+1$ & $-1$&$-1$\\[1ex] \hline \hline
\end{tabular}\label{Tab III}
\end{center}
\end{table}
\end{center}
\begin{center}
\begin{table}[t!]
\caption{ Eigenvalues of critical points in table \ref{Tab III} }
\begin{center}
\begin{tabular}{ccccc}
\hline\hline
~~~~Point~~~~& $~~~~E_1~~~~~~$ & $~~~~~~~~~E_2~~~~~~~~~$ & $~~~~~~E_3~~~~~~$&$~~~C_s^2~~~$
 \\ \hline\\
$B_{1}$ & $3-\frac{\sqrt{6\theta}}{2}\lambda$  & $\frac{3}{2}$ & $3\,(2-\gamma)-2\,\alpha\sqrt{\theta}$&$1$
 \\ [2ex]
$B_{2}$ & $3+\frac{\sqrt{6\theta}}{2}\lambda$  & $\frac{3}{2}$ & $3\,(2-\gamma)+2\,\alpha\sqrt{\theta}$&$1$
 \\ [2ex]
$B_3$ & $ \frac{1}{12}\,\frac{4\,\alpha^{2}\theta+9\,\gamma\,(2-\gamma)}{(2-\gamma)} $ &$ \frac{1}{6}\,\frac{4\,\alpha^{2}\theta-9\,(2-\gamma)^{2}}{(2-\gamma)} $ & $ \frac{1}{6}\,\frac{-2\sqrt{6}\lambda\alpha\theta+4\,\alpha^{2}\theta+9\,\gamma(2-\gamma)}{(2-\gamma)} $&$1$
 \\ [2ex]
$B_4$ & $\frac{\theta\lambda^2}{4}$ & $\frac{\theta\lambda^2}{2}-3$ & $-\frac{1}{3}\sqrt{6}\alpha\lambda\theta+\theta\lambda^2-3\gamma$&$\frac{\theta\,\lambda^2}{3}-1$
 \\ [2ex]
$B_5$ & $\frac{\gamma(3b+\sqrt{6}\alpha)}{4b}$ & $\mu_+$ & $\mu_-$ &$-1+\frac{9\,\theta\gamma^2}{9\theta\,\gamma-\sqrt{6}\alpha\,b}$
\\ [2ex]
$B_6$ & $0$ & $-3$ & $-3\,\gamma$&stable (limiting)
\\[1ex] \hline \hline\\
\end{tabular}\label{Tab IV}
\end{center}
where, $\mu_{\pm}=-\frac{\left(3b(2-\gamma)-\gamma\sqrt{6}\alpha\right)}{4b}\left[1\pm\sqrt{1-\frac{24\left(3\gamma(2-\gamma)-\frac{2}{3}\sqrt{6}\theta\alpha b \right)\left(\theta(\frac{\alpha\sqrt{6}b}{3}+b^2)-3\gamma\right)}{(3b(2-\gamma)-\gamma\sqrt{6}\alpha)^2}}\,\right]$
\end{table}
\end{center}
\begin{figure}[h!]
\centering
\subfigure[]{%
\includegraphics[width=8cm,height=6cm]{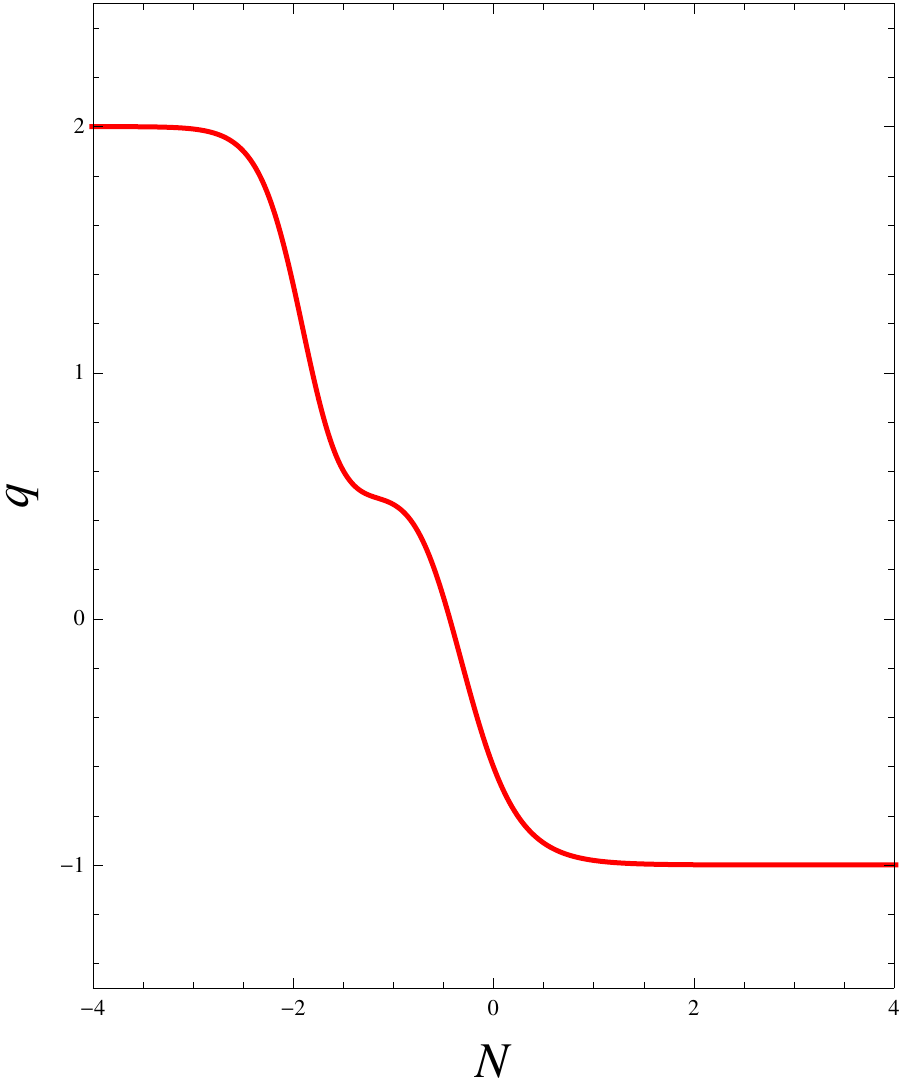}\label{decbq}}
\qquad
\subfigure[]{%
\includegraphics[width=8cm,height=6cm]{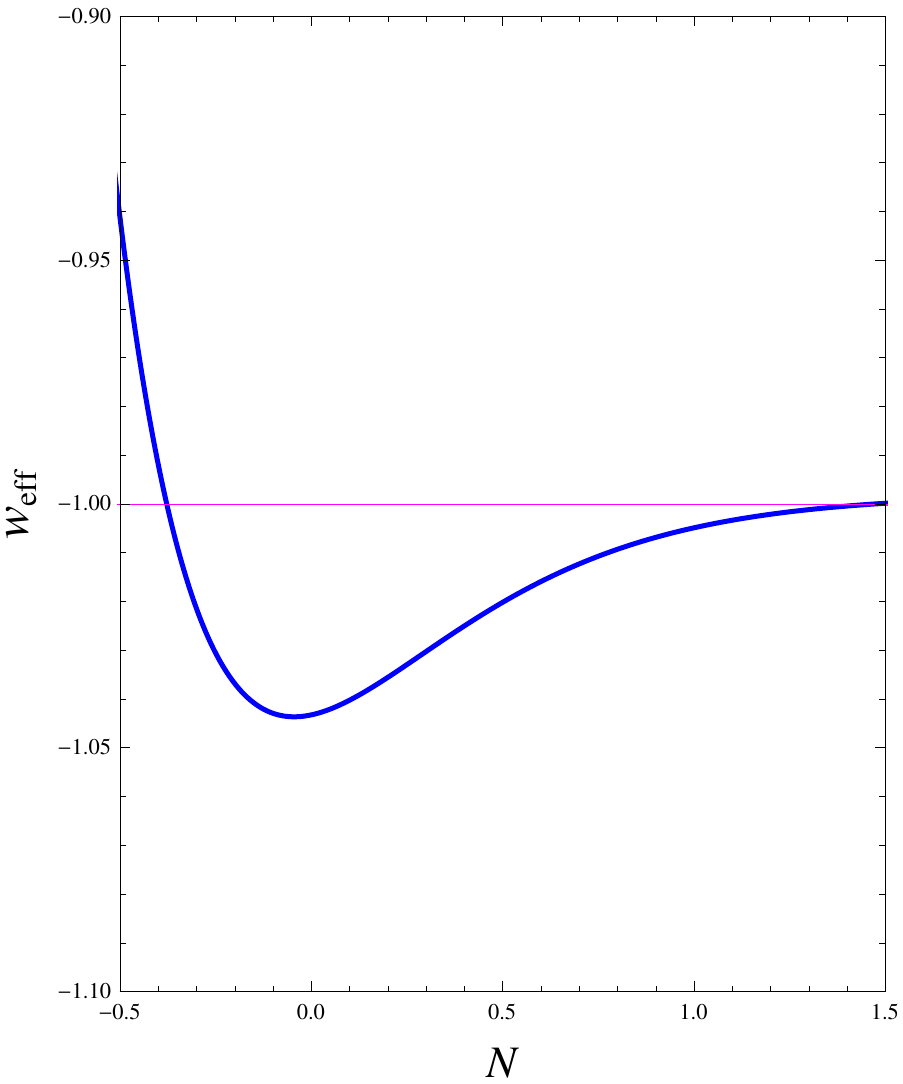}\label{weffbq}}
\caption{(a).  Plot of $q$ versus $N$. (b). Plot of $\omega_{\rm eff}$
\textit{vs N} for $\theta=1$, for $\Gamma=1$ with $\alpha=-0.7$,
$\gamma=1$.}
\end{figure}
\par In this case we see that universe evolves from one of the unstable points $B_1$ or $B_2$ and approaches toward any of the saddle points $B_3$, $B_4$ or $B_5$ and finally settles down towards the attracting set $B_6$. Furthermore, the attractor set $B_6$ is also classically stable and is very interesting from cosmological point of view.
\par
\begin{figure}[h!]
\centering
\subfigure[]{%
\includegraphics[width=8cm,height=6cm]{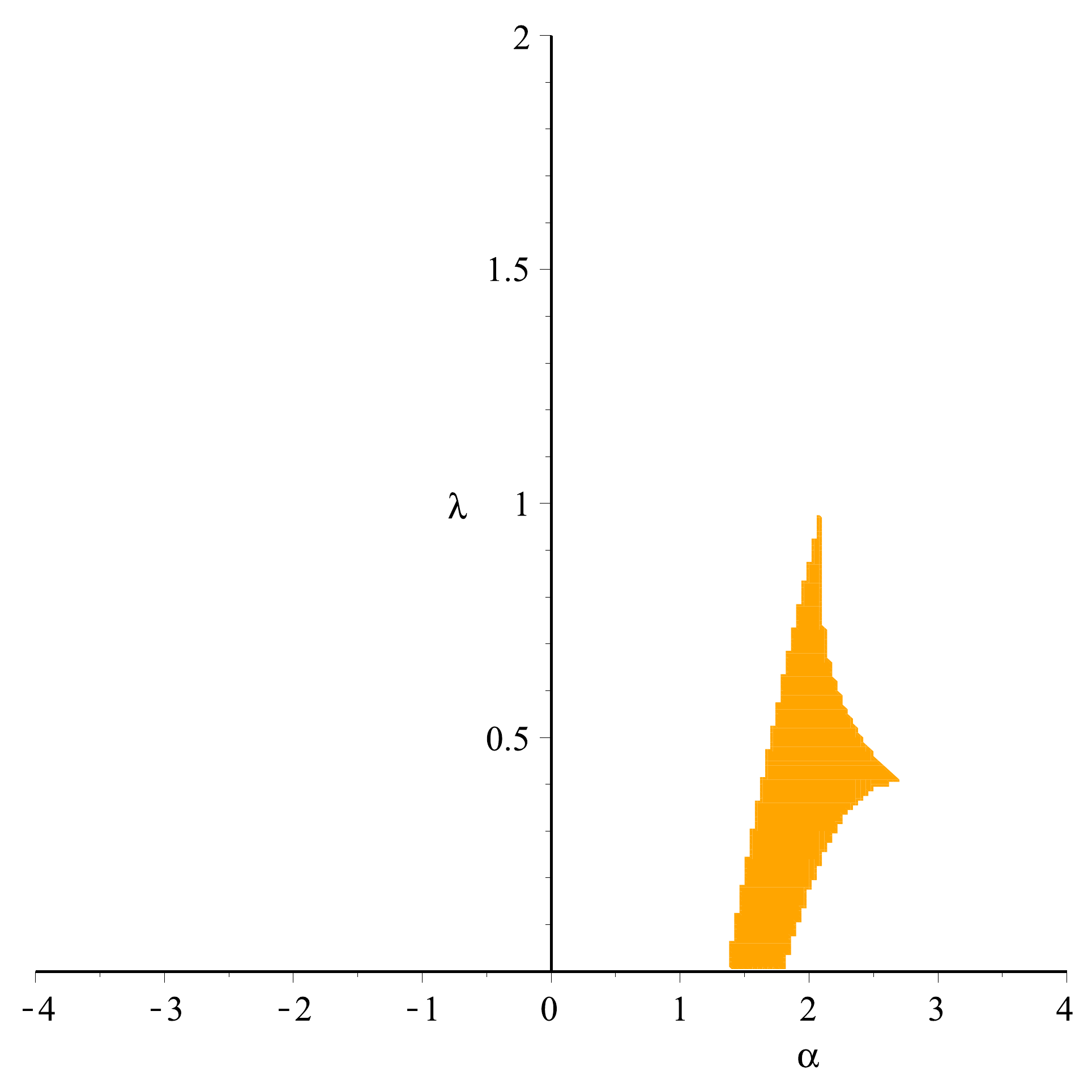}\label{b5}}
\qquad
\subfigure[]{%
\includegraphics[width=8cm,height=6cm]{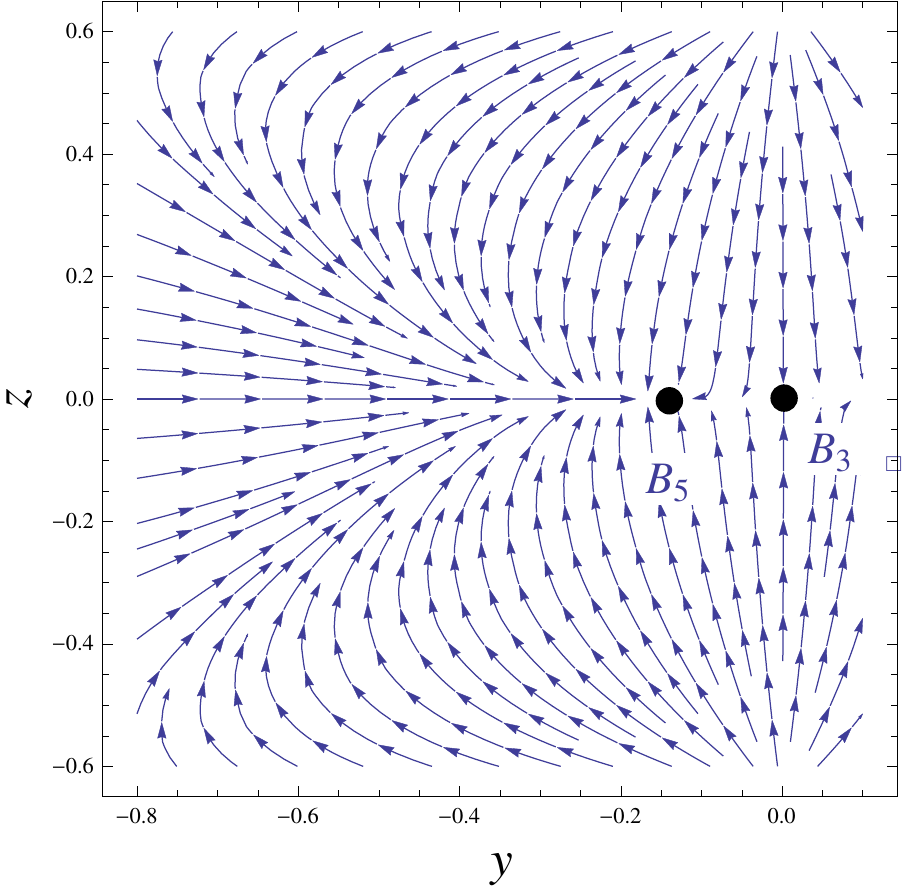}\label{spbeb5}}
\caption{(a). Region of stability of $B_5$. (b). $y-z$ plane projection of the system (\ref{DGP I:Sec3:17})-(\ref{DGP I:Sec3:19}) for $\theta=-1$ with $\alpha=1.8$, $\gamma=0.5$.}
\end{figure}

\begin{figure}
\centering
\includegraphics[width=8cm,height=6cm]{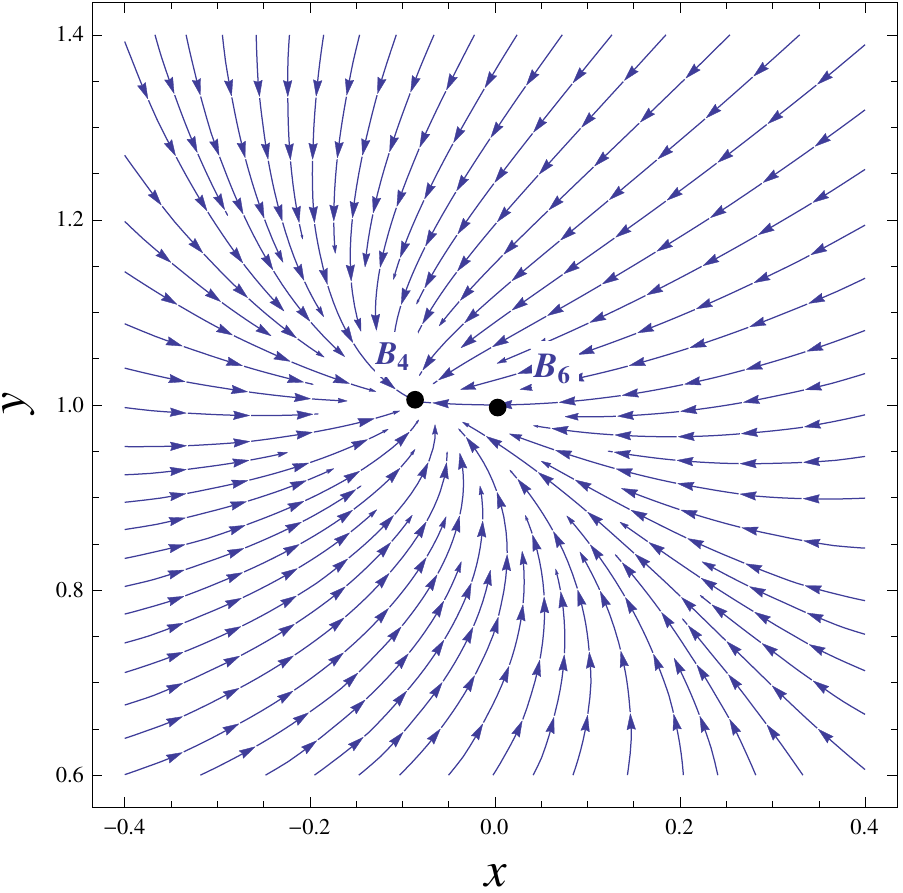}
\caption{ $x-y$ plane projection of the system (\ref{DGP I:Sec3:17})-(\ref{DGP I:Sec3:19}) for $\theta=-1$.}\label{spbe}
\end{figure}

 The behaviour of deceleration parameter $q$ for the case of $\Gamma=1$ is given in fig.\ref{decbq}. The universe undergoes transition from decelerated phase to an accelerated phase around $N=-0.44$ (equivalent to a redshift of $0.55$). This indeed matches with the observation \cite{Liang}.
  Finally universe settles down with an accelerated expansion ($q=-1$).  Also crossing of phantom divide is possible as shown in fig.\ref{weffbq} for quintessence field. A similar behaviour can be observed for the case of $\Gamma\neq 1$ also. The crossing of phantom divide line can also be understood analytically. For $\omega_{\phi} > -1$ and $\gamma = 1$, we see from eqn.(\ref{DGP I:Sec02:08b}) that 1+$\omega_{\rm eff}$ can assume positive as well as negative values. Hence $\omega_{\rm{eff}}$ can pass through -1.\\\linebreak
(ii) Phantom field ($\theta=-1$):\\
Critical points $B_1,\, B_2$ do not exist in this case.  Point $B_3$ is physically meaningless since $\Omega_{\phi}$ is negative. Point $B_4$ corresponds
 to an accelerated phantom field dominated solution ($\Omega_{\phi}=1,\, q=-\frac{\lambda^2}{2}-1$).
It is a late time attractor if $\alpha<\frac{\sqrt{6}}{2\,\lambda}(\lambda^2+3\gamma)$. However this point is not classically stable. Scaling solution $B_5$ is stable for a narrow
range of parameters. Fig.\ref{b5} shows the region of stability and the complicated conditions for stability is confirmed numerically. It can be seen that, if we numerically put $\alpha=1.8,\,\lambda=0.7,\,\gamma=0.5$, we obtain $E_1=-0.34,\,E_2=-3.67,\,E_3=-0.008,\, C_s^2=1.03,\, q=-1.68,\,\Omega_{\phi}=0.64$ within the region of existence of $B_5$. Thus we get a late time accelerated scaling attractor in this case. Fig. \ref{spbeb5} shows the projection of the system on $y-z$ plane. This is indeed an  interesting point since this point  is not obtained in case of corresponding uncoupled DGP model \cite{Nozari}.  For the non isolated set of critical points $B_6$, we plot a projection of a system on the $x-y$ plane (fig.\ref{spbe}). We observe that a point $(0,1)$ which lies on $x-y$ plane is unstable and hence $B_6$ is unstable. In this case universe starts from an unknown point and finally settles down to point $B_5$.
\par
\begin{figure}[h!]
\centering
\subfigure[]{%
\includegraphics[width=8cm,height=6cm]{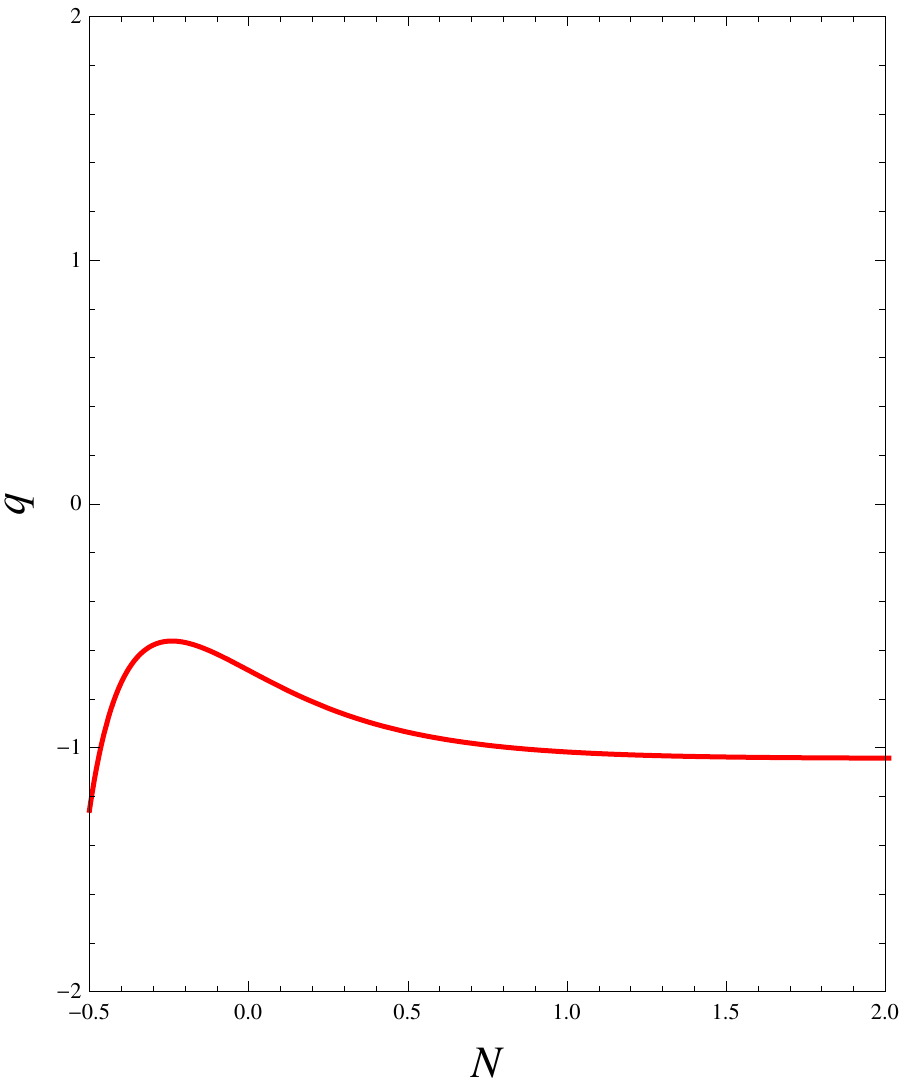}\label{decbp}}
\qquad
\subfigure[]{%
\includegraphics[width=8cm,height=6cm]{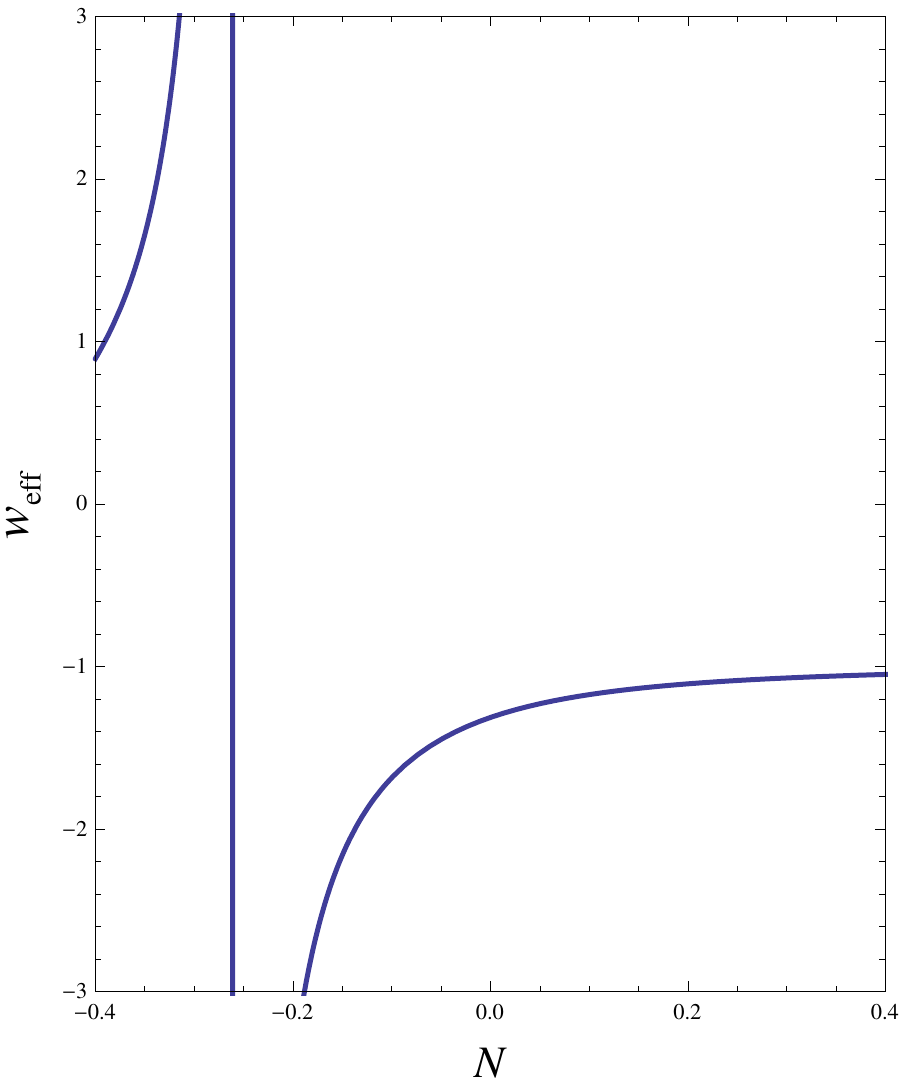}\label{weffbp}}
\caption{(a). Plot of $q$ versus $N$  (b). Plot of $\omega_{\rm eff}$
\textit{vs N} for $\theta=-1$, $\Gamma=1$ with $\alpha=-0.7$,
$\gamma=1$.}
\end{figure}

 The behaviour of deceleration parameter $q$ for the case of $\Gamma=1$ is given in fig.\ref{decbp}. The universe is always in accelerated phase.
Also crossing of phantom divide is not possible as shown in fig.\ref{weffbp}. A similar behaviour can be observed for the case of $\Gamma\neq 1$ also. This also can be understood analytically, since for $\omega_{\phi}<-1$ and $\gamma=1$, the numerator of eqn.(\ref{DGP I:Sec02:08b}) is always positive. In fig.\ref{weffbp}, $\omega_{\rm eff}$ seems to diverge. Actually no pathology is associated to the model and this divergence is associated with effective behaviour.  This happens because $\Omega_{\rm eff}$ evolve from either positive to negative values or vice versa. So, $\Omega_{\rm eff}=0$ at some values of $N$, which leads to the breakdown of the effective behaviour. This sort of behaviour is also obtained in \cite{Chimento,Wu}.

\subsection{\textbf{Interaction B: $Q=\beta\,\dot{\rho}_{\phi}$}}\label{intII}
\noindent The autonomous system of equation for this interaction is given by
\begin{eqnarray}
\label{DGP I:Sec3:22} x'&=& \frac{3x}{(\beta-1)}+\theta \frac{1}{2}\,s\,\sqrt {6}\,{y}^{2}+\frac{3}{2}\,{\frac {x \left( 2 \theta\,{x}^{2}+\gamma\, \left( 2\,{z}^{2}-\theta{x}^{2}-{y}^{2}+1 \right)  \right) }{{z}^{2}+1}}\\
\label{DGP I:Sec3:23}y' &=& -\frac{1}{2}\,s\sqrt {6}\,x\,y+\frac{3}{2}\,{\frac {y \left( 2\theta \,{x}^{2}+\gamma\, \left( 2\,{z}^{2}-\theta{x}^{2}-{y}^{2}+1 \right)  \right) }{{z}^{2}+1}} \\
\label{DGP I:Sec3:24} z'&=& \frac{3}{4}\,{\frac {z \left(2 \theta\,{x}^{2}+\gamma\, \left( 2\,{z}^{2}-\theta{x}^{2}-{y
}^{2}+1 \right)  \right) }{{z}^{2}+1}}\\
\label{DGP I:Sec3:25} s'&=& -\sqrt{6}\,x\,{s}^{2}\left(\Gamma-1\right)
\end{eqnarray}
It is noted when $x=0$, the first term on the right hand side of eq.(\ref{DGP I:Sec3:22}) containing $\beta$ vanishes. So, point where $x=0$ is independent of the interaction $Q$ and hence exists in uncoupled model also. It can also be seen that the system is invariant  under the change of sign $y\rightarrow -y$ and $z\rightarrow -z$, so we restrict our analysis to the positive values of $y$ and $z$.\\\linebreak
The adiabatic speed of sound $C_s^2$ is given by
\begin{equation}
1+\frac{2 \theta (\beta-1)s\,y^2}{\sqrt{6}x}\label{DGP I:Sec3:26}
\end{equation}
As before, in what follows we study the phase space analysis for the two categories of potentials.
\subsubsection{\textbf{Category I:}\textit{Non-exponential form of potential} ($\Gamma\neq1$)} \label{sub3,3}
\noindent In this category eqs.(\ref{DGP I:Sec3:22})-(\ref{DGP I:Sec3:25}) form a closed system of equations. The critical points and their cosmological parameters are listed in table \ref{Tab V} and the eigenvalues of their corresponding Jacobian matrix are given in table \ref{Tab VI}. Like previous case, we discuss the stability of critical points for the two fields separately.
\begin{center}
\begin{table}[h!]
\caption[crit]{Critical points and their cosmological parameters of system. We have defined: $\xi_{\pm}=\pm\sqrt{-\frac{\left(\beta\gamma+(2-\gamma)\right)}{\theta(\beta-1)(2-\gamma)}}$}
\begin{center}
\begin{tabular}{ccccccccc}
\hline\hline
~~~Point~~~&$~~~x~~~$&$~~~y~~~$&$~~~z~~~$
&$~~~s~~~$&~~~Existence~~~&$~~~\Omega_\phi~~~$&$~~~\omega_\phi~~~$&$~~~q~~~$ \\ \hline\\[0ex]
$C_{1}$&$0$&$0$&$0$&$s$&Always  & $0$ & $1$&$-1+\frac{3\gamma}{2}$ \\ [2ex]
$C_{2}$&$\xi_{+}$&$0$&$0$&$0$&\raisebox{2ex}{$0<\theta\xi_{+}^2<1$}  & $\theta\xi_{+}^2$ & $1$&$-\frac{\beta+2}{\beta-1}$ \\
&&&&&\raisebox{2ex}{$\beta\neq 1$, $\gamma\neq 2$}&&&\\
$C_{3}$&$\xi_{-}$&$0$&$0$&$0$&\raisebox{1.5ex}{$0<\theta\xi_{-}^2<1$}  & $\theta\xi_{-}^2$ & $1$&$-\frac{\beta+2}{\beta-1}$ \\
&&&&&\raisebox{2ex}{$\beta\neq 1$, $\gamma\neq 2$}&&&\\
$C_4$ & $0$&$\sqrt{2\,z^{2}+1}$&$z$&$0$&Always  & $2\,z^{2}+1$ & $-1$&$-1$\\[1ex] \hline \hline
\end{tabular} \label{Tab V}
\end{center}
\end{table}
\end{center}
\begin{center}
\begin{table}[h!]
\caption{ Eigenvalues of critical points in table \ref{Tab V} }
\begin{center}
\begin{tabular}{cccccc}
\hline\hline
~~~~~Point~~~~~& $~~~~E_1~~~~$ & $~~~~~~E_2~~~~~~$ & $~~~~E_3~~~~$ & $~~~~E_4~~~~$&$~~~C_s^2$
\\ \hline\\
$C_{1}$ & $\frac{3}{\beta-1}+\frac{3\gamma}{2}$ & $\frac{3\gamma}{2}$ & $\frac{3\gamma}{4}$ & $0$& stable (limiting)
 \\ [2ex]
$C_{2}$ & $3\theta(2-\gamma)\,\xi_{+}^2$ & $-\frac{3}{\beta-1}$ & $-\frac{3}{2(\beta-1)}$ & $0$&$1$
\\ [2ex]
$C_{3}$ & $3\theta(2-\gamma)\,\xi_{-}^2$ & $-\frac{3}{\beta-1}$ & $-\frac{3}{2(\beta-1)}$ & $0$&$1$
\\ [2ex]
$C_4$ & $-3\gamma$ & $\frac{3}{\beta-1}$ & $0$ & $0$& undefined
\\[1ex] \hline\hline
\end{tabular} \label{Tab VI}
\end{center}
\end{table}
\end{center}
\begin{figure}[h!]
\centering
\subfigure[]{%
\includegraphics[width=8cm,height=6cm]{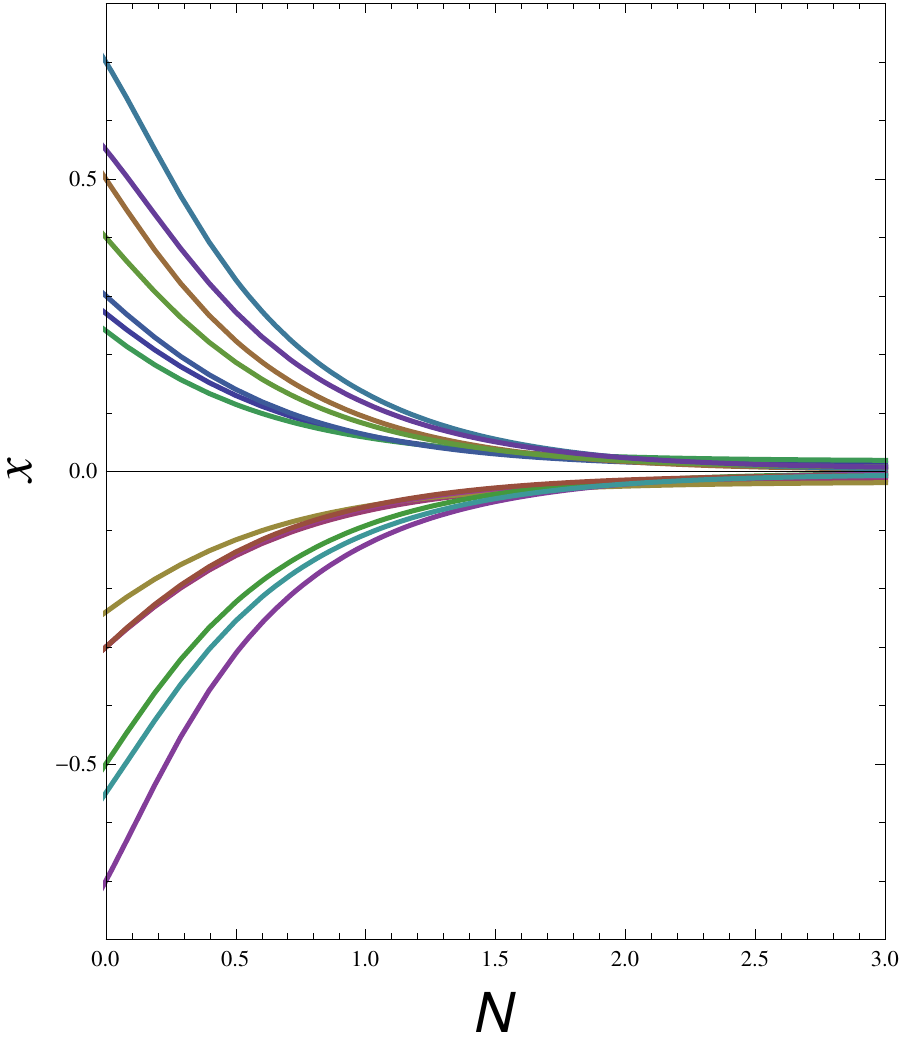}\label{fig6}}
\qquad
\subfigure[]{%
\includegraphics[width=8cm,height=6cm]{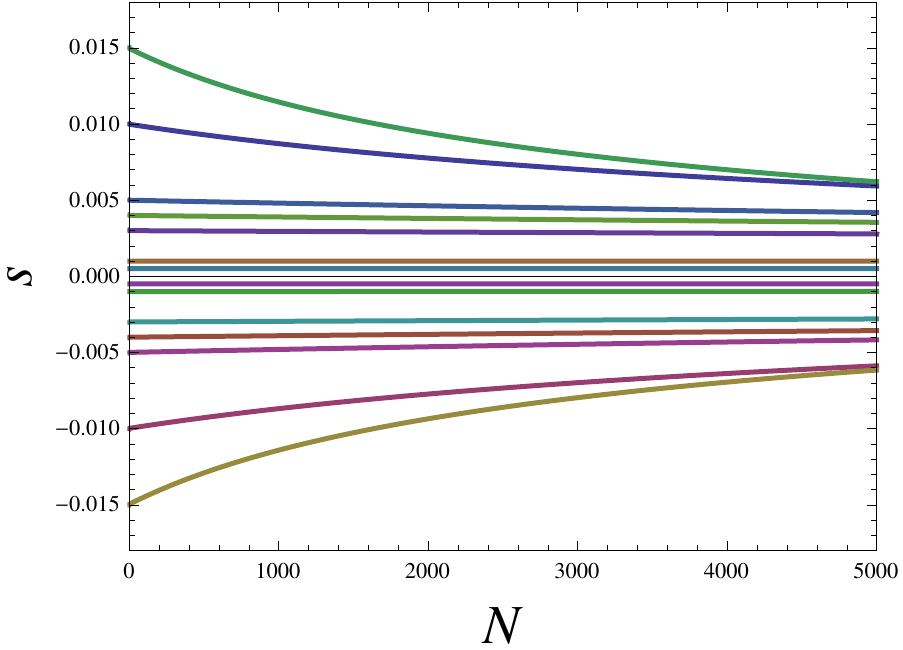}\label{fig9}}\\
\qquad
\subfigure[]{%
\includegraphics[width=8cm,height=6cm]{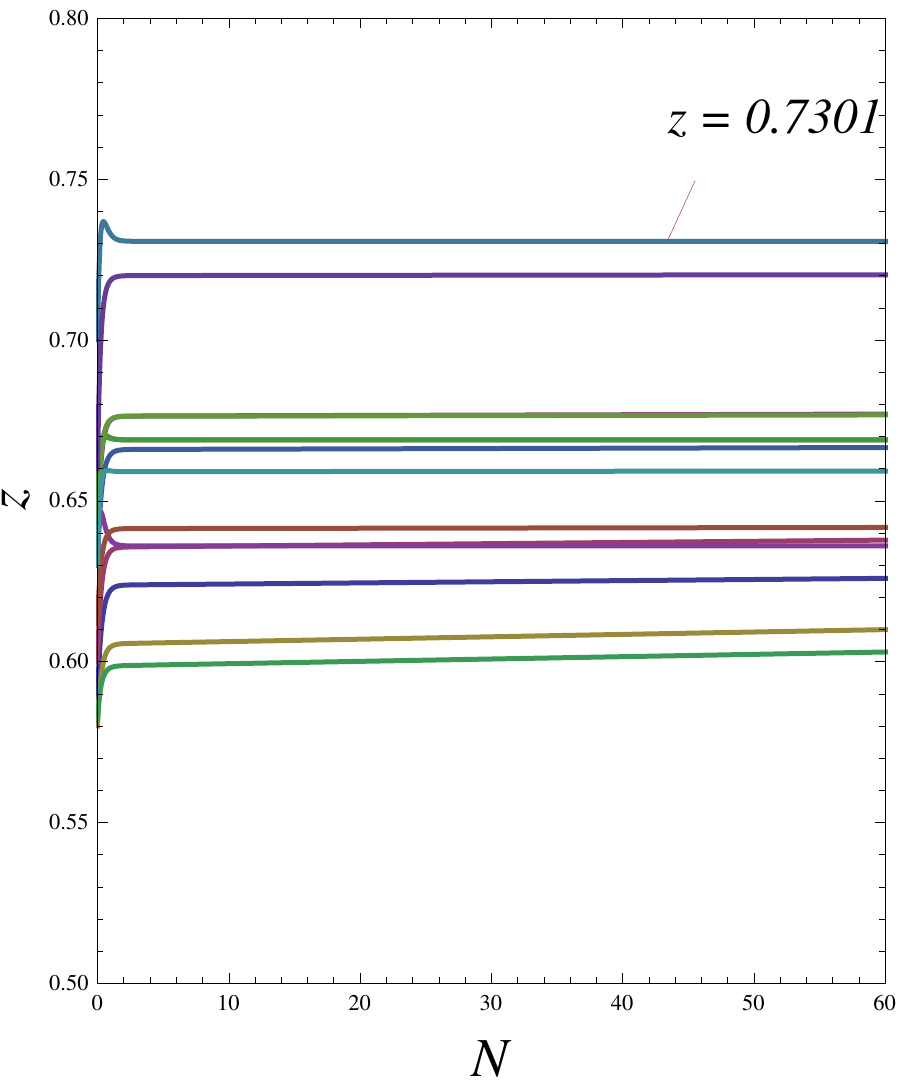}\label{fig8}}
\qquad
\subfigure[]{%
\includegraphics[width=8cm,height=6cm]{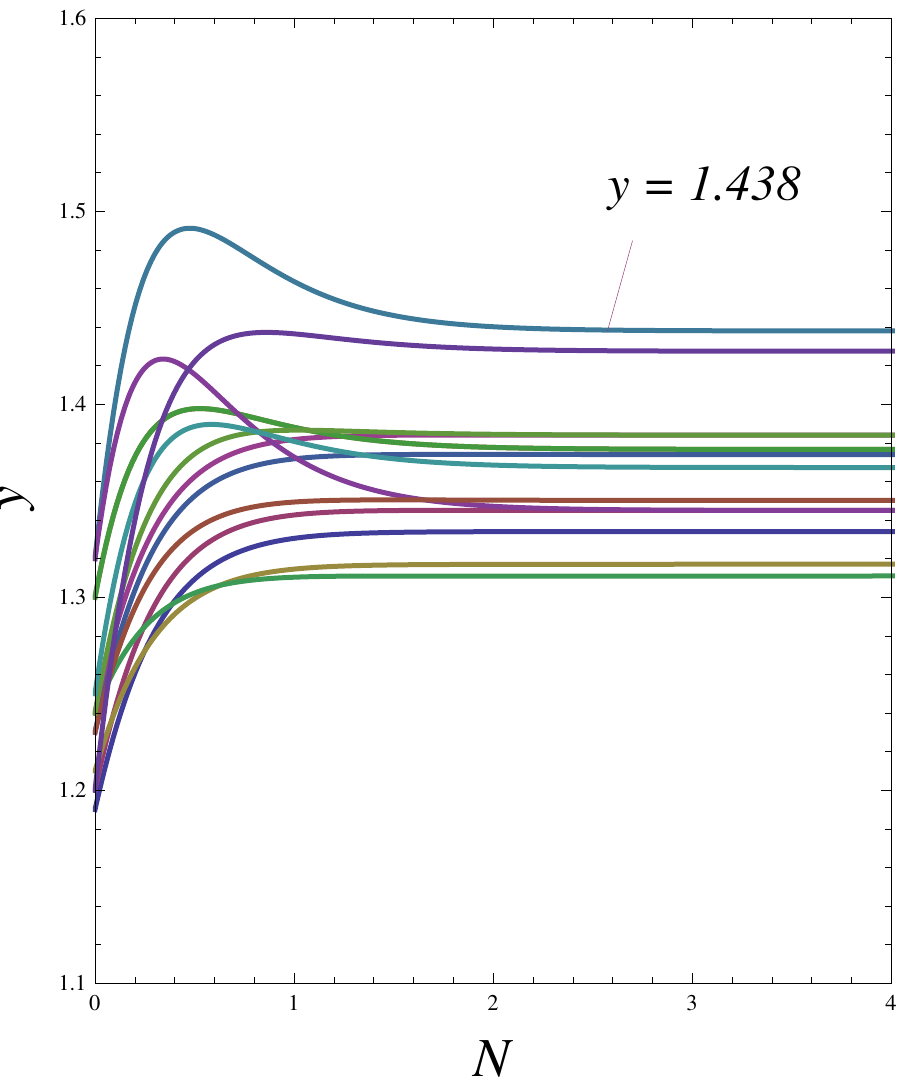}\label{fig7}}

\caption{(a). Projection of perturbation plot of $x$ versus $N$.  (b). Projection of perturbation plot of $s$ versus $N$. (c). Projection of perturbation plot of $z$ versus $N$. (d). Projection of perturbation plot of $y$ versus $N$ for $\theta=1$.}
\end{figure}
(i) Quintessence field ($\theta=1$):\\
$C_1$  corresponds to an accelerated matter dominated universe ($\Omega_m=1$) for $\gamma<\frac{2}{3}$. It is an unstable node for $\frac{\gamma}{2}>\frac{1}{1-\beta}$
otherwise it behaves as a saddle point. Scaling solutions $C_2$ and $C_3$ are unstable nodes for $\beta<1$ else they behave as saddle points. Set of critical points $C_4$
corresponds to an accelerated solution ($q=-1$). Since two of its eigenvalues are zero and the other two are negative provided $\beta<1$, (it behaves as a saddle point for $\beta>1$),
so linear stability theory is not enough and further investigation is required . To check the stability of this non isolated set, we numerically perturb the solutions around
 the critical point. We again plot the projections plots on $x,y,z$ and $s$ separately. Like previous case, from figs.\ref{fig6}-\ref{fig7}, it is evident that the non isolated
 set of critical points $C_4$ is a late time attractor.
\par  In this case, universe evolves from one of these unstable points $C_2$ or $C_3$ and approaches toward the saddle point $C_1$ and finally settles down towards the attractor set $C_4$.
\\\linebreak
(ii) Phantom field ($\theta=-1$):\\
Critical points $C_2,\,C_3$  do not exist for phantom field. Critical point $C_1$ is unstable node for $\frac{1}{(1-\beta)}<\frac{\gamma}{2}$,
 otherwise it behaves as saddle point.  The set of critical points $C_4$ corresponds to an accelerated solution. Since its corresponding Jacobian matrix contains two zero and two negative
 eigenvalues, further investigation is required. As in case of interaction \textbf{A}, we plot a 2D projection of the system on the $x-s$ plane  and we observe
  that trajectories which initially approach a point $(0,0)$ which lies on set $C_4$, ultimately diverge away from it. This implies that a set of points $C_4$ is an unstable
  set. Thus we do not get any interesting cosmological scenario in
  this case.
\subsubsection{\textbf{Category II}: \textit{Exponential form of potential} ($\Gamma=1$)} \label{sub3,4}
\noindent In this category, $s$ is constant and $V=V_0\,\rm exp
(-\lambda\phi)$, so eqs.(\ref{DGP I:Sec3:22})-(\ref{DGP
I:Sec3:24}) form a closed system of equations. Critical points and
corresponding cosmological parameters are listed in table \ref{Tab
VII} and the eigenvalues of their corresponding Jacobian matrix
are given in table \ref{Tab VIII}. As before, we discuss the
stability of critical points for two scalar fields separately.
\begin{center}
\begin{table}[h!]
\caption[crit]{Critical points and cosmological parameters. We have defined: $\xi_{\pm}=\pm\sqrt{-\frac{\left(\beta\gamma+(2-\gamma)\right)}{\theta(\beta-1)(2-\gamma)}}$}
\begin{center}
\begin{tabular}{cccccccc}
\hline\hline
~~~Point~~~&$~~~x~~~$&$~~~y~~~$&$~~~z~~~$&~~~~~Existence~~~~~&$~~~~~\Omega_\phi~~~~~$&$~~~~~\omega_\phi~~~~~$&$q$\\ \hline\\
$D_{1}$&$0$&$0$&$0$&Always  & $0$ & $1$&$-1+\frac{3\gamma}{2}$ \\ [2ex]
$D_{2}$&$\xi_{+}$&$0$&$0$&$0<\theta\xi_{+}^2<1$  & $\theta\xi_{+}^2$ & $1$&$-\frac{\beta+2}{\beta-1}$ \\ [2ex]
$D_{3}$&$\xi_{-}$&$0$&$0$&$0<\theta\xi_{-}^2<1$  & $\theta\xi_{-}^2$ & $1$&$-\frac{\beta+2}{\beta-1}$ \\ [2ex]
$D_4$ & $0$&$\sqrt{2\,z^{2}+1}$&$z$&Always  & $2\,z^{2}+1$ & $-1$&$-1$\\ [2ex]
$D_5$&$x_5$&$y_5$&
$0$&$\theta\,x_5^2+y_5^2<1$ & $x_5^2+y_5^2$ & $\frac{x_5^2-y_5^2}{x_5^2+y_5^2}$&$\frac{1}{16\lambda^2(\beta-1)^2}\left[\left\lbrace\lambda^2(\beta-1)+10\beta+2\delta-7\right\rbrace^2\right.$\\&&&&&&&$\left.-27+3\delta^2-(10\beta+2\delta-7)^2\right]$ \\\hline \hline\\\linebreak
\end{tabular} \label{Tab VII}
\end{center}
For quintessence field,\\
$x_5=\frac {\sqrt {6} \left( {\lambda}^{2} \left( \beta-1 \right) -3+\delta \right) }{12\,\lambda\, \left( \beta-1 \right) },\,
y_5={\frac {\sqrt {-3\,({\delta}^{2}+6\,\beta \left( {\lambda}^{2}\left( \beta-1 \right) +3 \right)-18\,(\beta+1)+\left( {\lambda}^{2} \left( \beta-1 \right) +3 \right) \delta)}}{6\,\lambda\, \left( \beta-1 \right) }},\,
\delta=\sqrt{{\lambda}^{2} \left( \beta-1 \right)  \left( {\lambda}^{2}(\beta\,-1)+6(1-2\beta) \right)+9} $\\
For phantom field,\\
$x_5={\frac {\sqrt {6} \left(-{\lambda}^{2} \left( \beta-1\right) -3+\delta \right) }{12\lambda\, \left( \beta-1 \right) }},\,
 y_5=\frac {\sqrt {3\,({\delta}^{2}-6\,\beta \left( {\lambda}^{2} \left( \beta-1 \right) -3 \right) -18(\beta+1)+\, \left( {\lambda}^{2} \left( \beta-1 \right) -3 \right) \delta)}}{6\lambda\, \left( \beta-1 \right) },\,
 \delta=\sqrt{{\lambda}^{2} \left( \beta-1 \right)  \left( {\lambda}^{2}(\beta\,-1)-6(1-2\beta) \right)+9} $
\end{table}
\end{center}
\begin{figure}[h!]
\centering
\subfigure[]{%
\includegraphics[width=7cm,height=5cm]{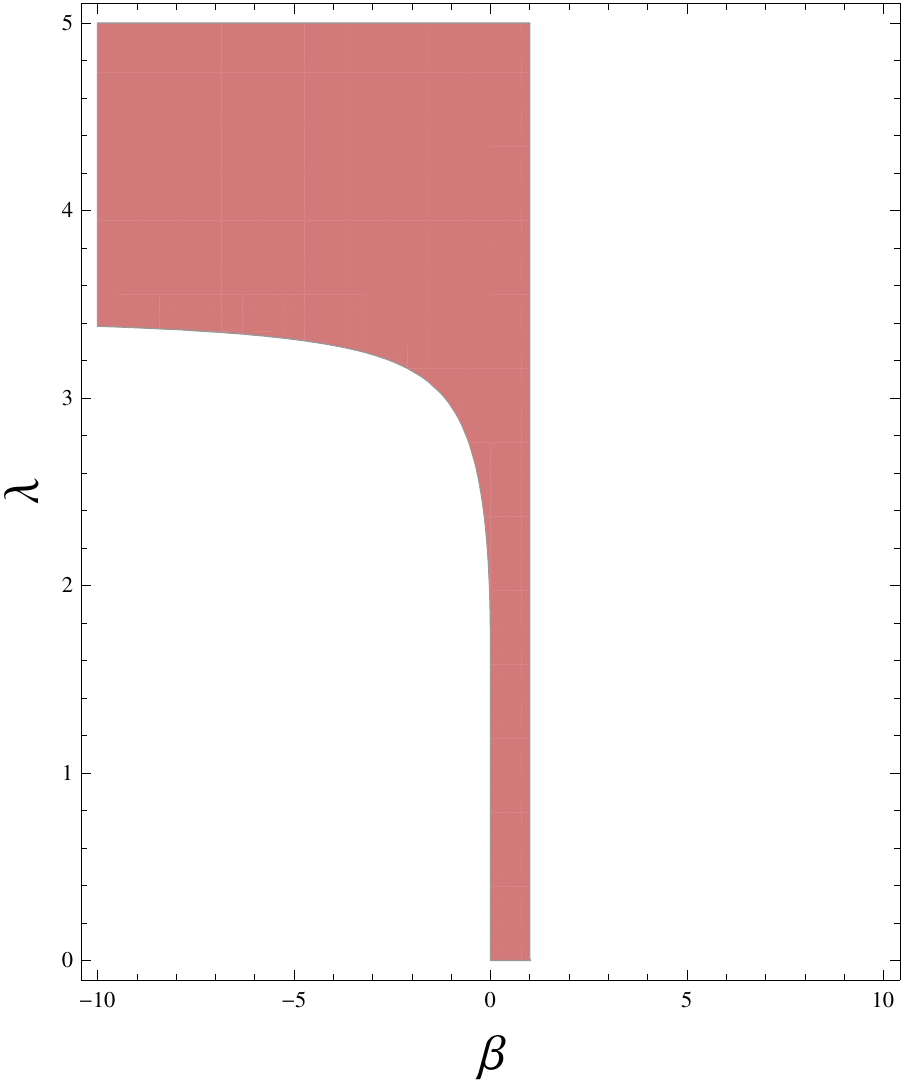}\label{fig11}}
\qquad
\subfigure[]{%
\includegraphics[width=7cm,height=5cm]{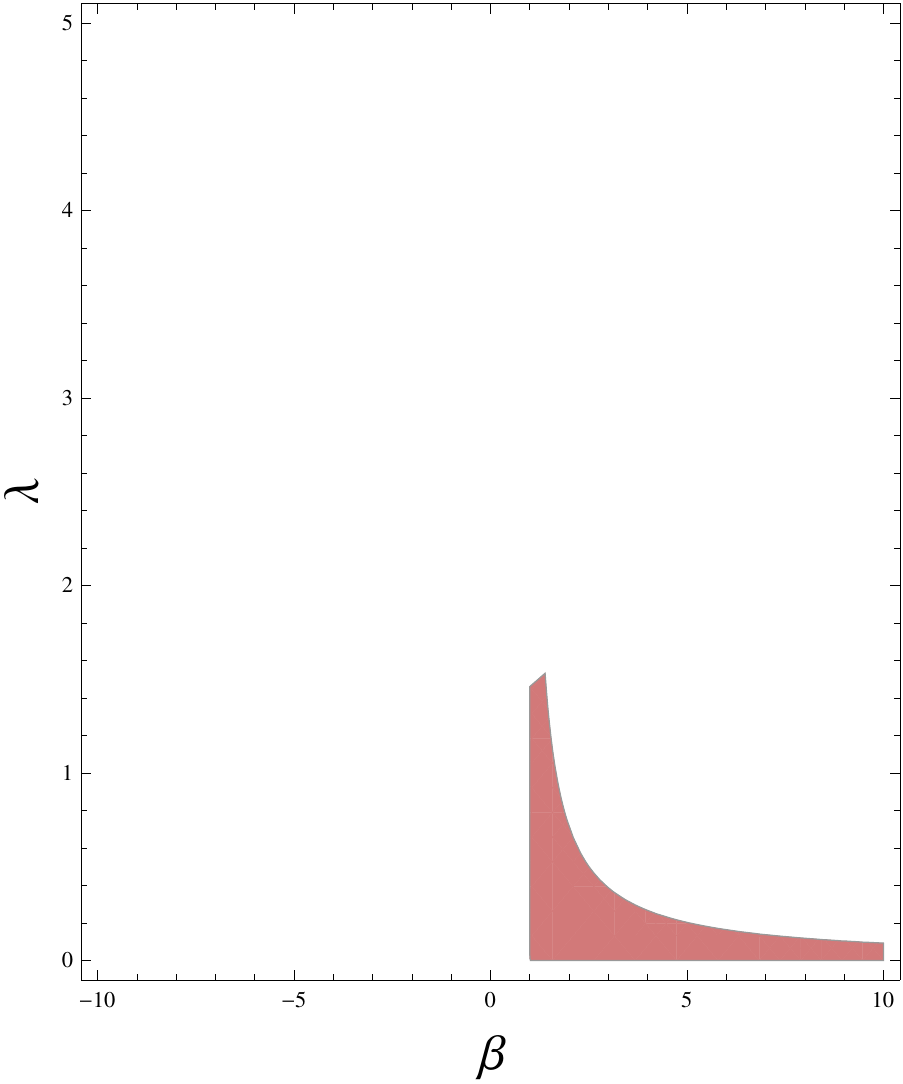}\label{fig12}}
\caption{(a). Existence region of point $D_5$. (b). Region for
negativity of one eigenvalue $\eta_1$ with $\theta=1$.}
\end{figure}
\begin{figure}[h!]
\centering
\subfigure[]{%
\includegraphics[width=7cm,height=5cm]{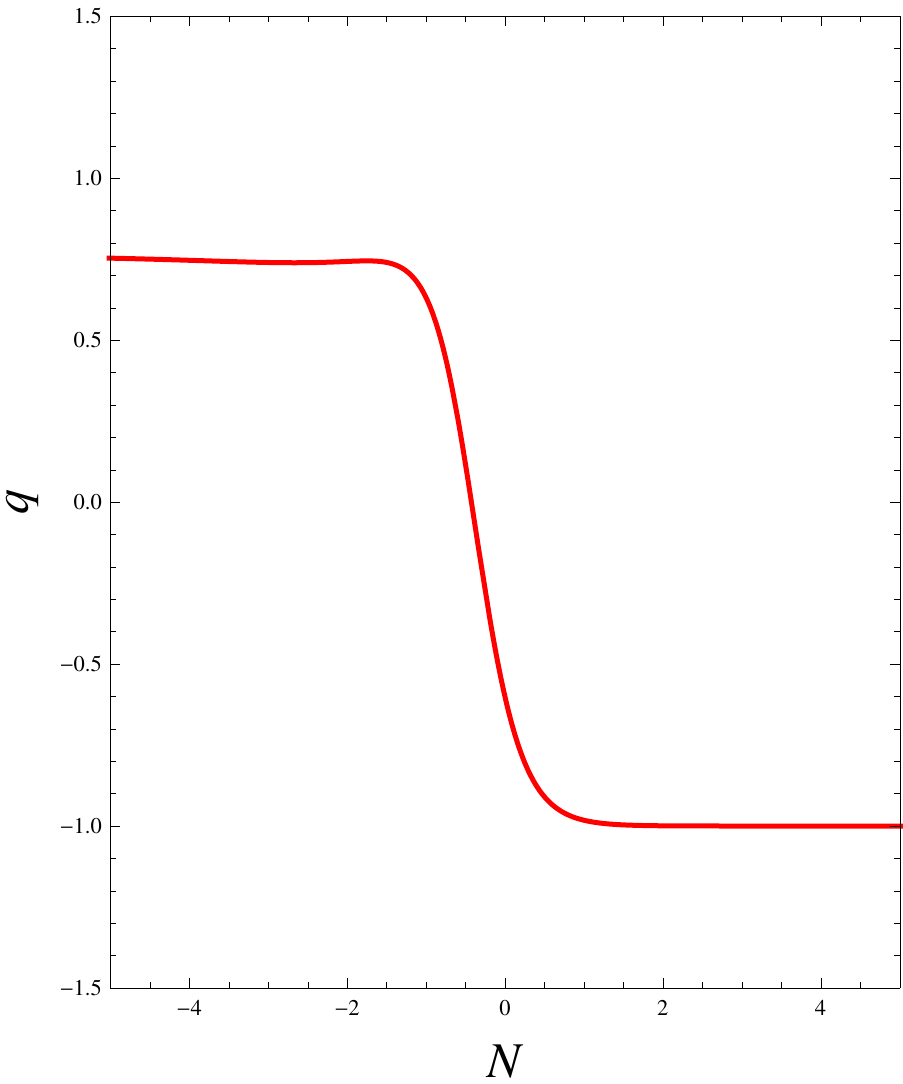}\label{decpq}}
\qquad
\subfigure[]{%
\includegraphics[width=7cm,height=5cm]{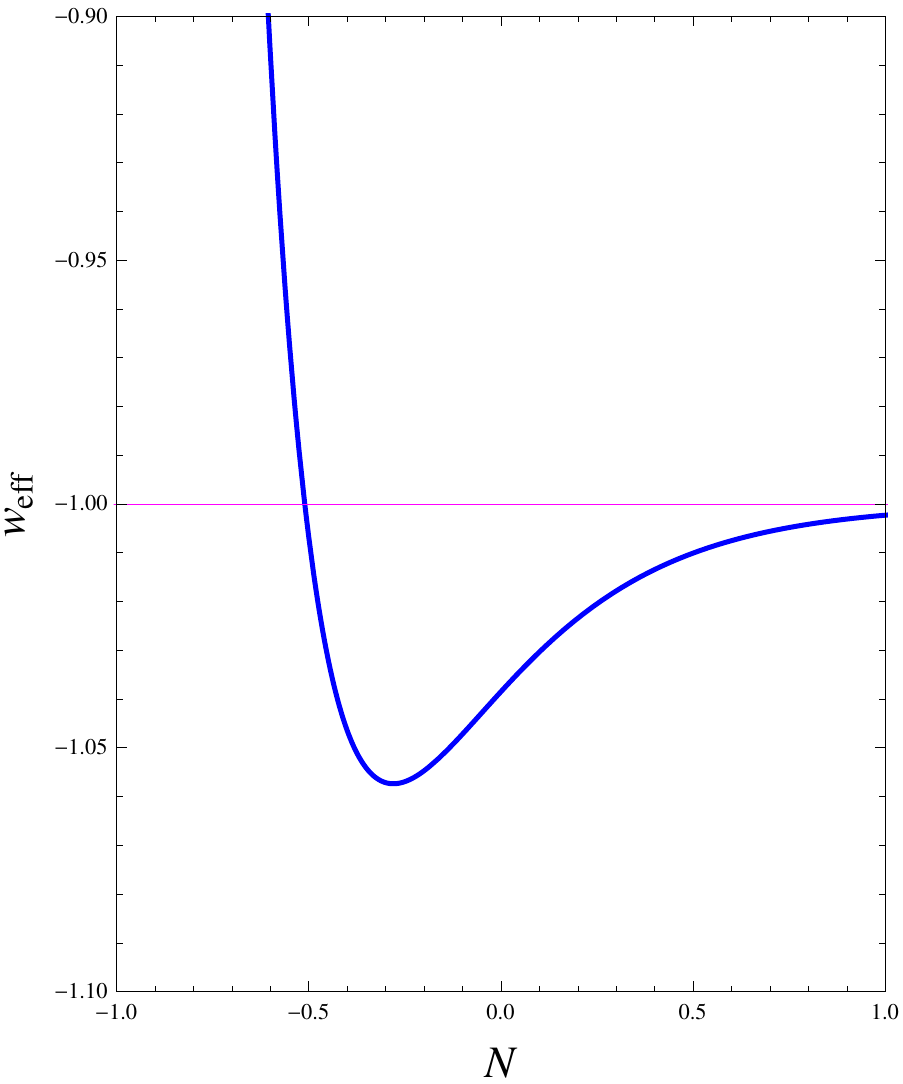}\label{weffpq}}
\caption{(a). Plot of $q$ versus $N$. (b). Plot of $\omega_{\rm eff}$ \textit{vs N}  for  $\theta=1$, $\Gamma=1$ with $\beta=-0.7$, $\gamma=1$.}
\end{figure}
\begin{figure}[h!]
\centering
\includegraphics[width=7cm,height=5cm]{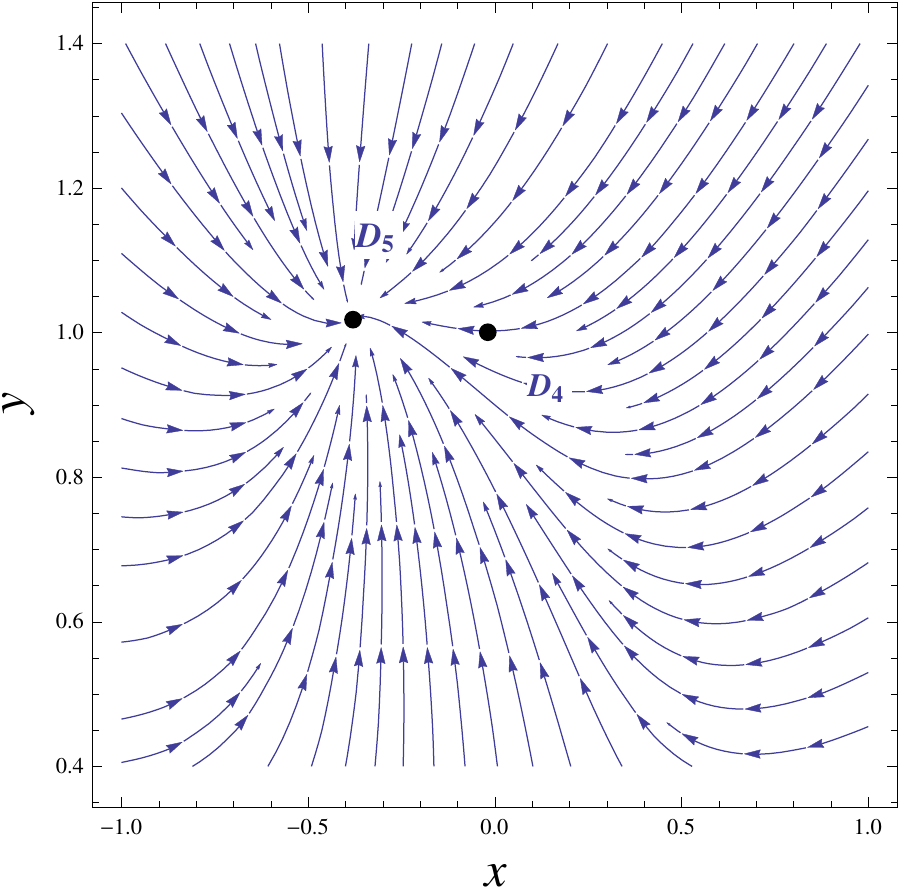}
\caption{$x-y$ plane projection of the system (\ref{DGP
I:Sec3:22})-(\ref{DGP I:Sec3:24}) for $\theta=-1$. It seems that
point $D_5$ is stable but actually not stable.}\label{sppe}
\end{figure}
\begin{figure}[t!]
\centering
\subfigure[]{%
\includegraphics[width=8cm,height=6cm]{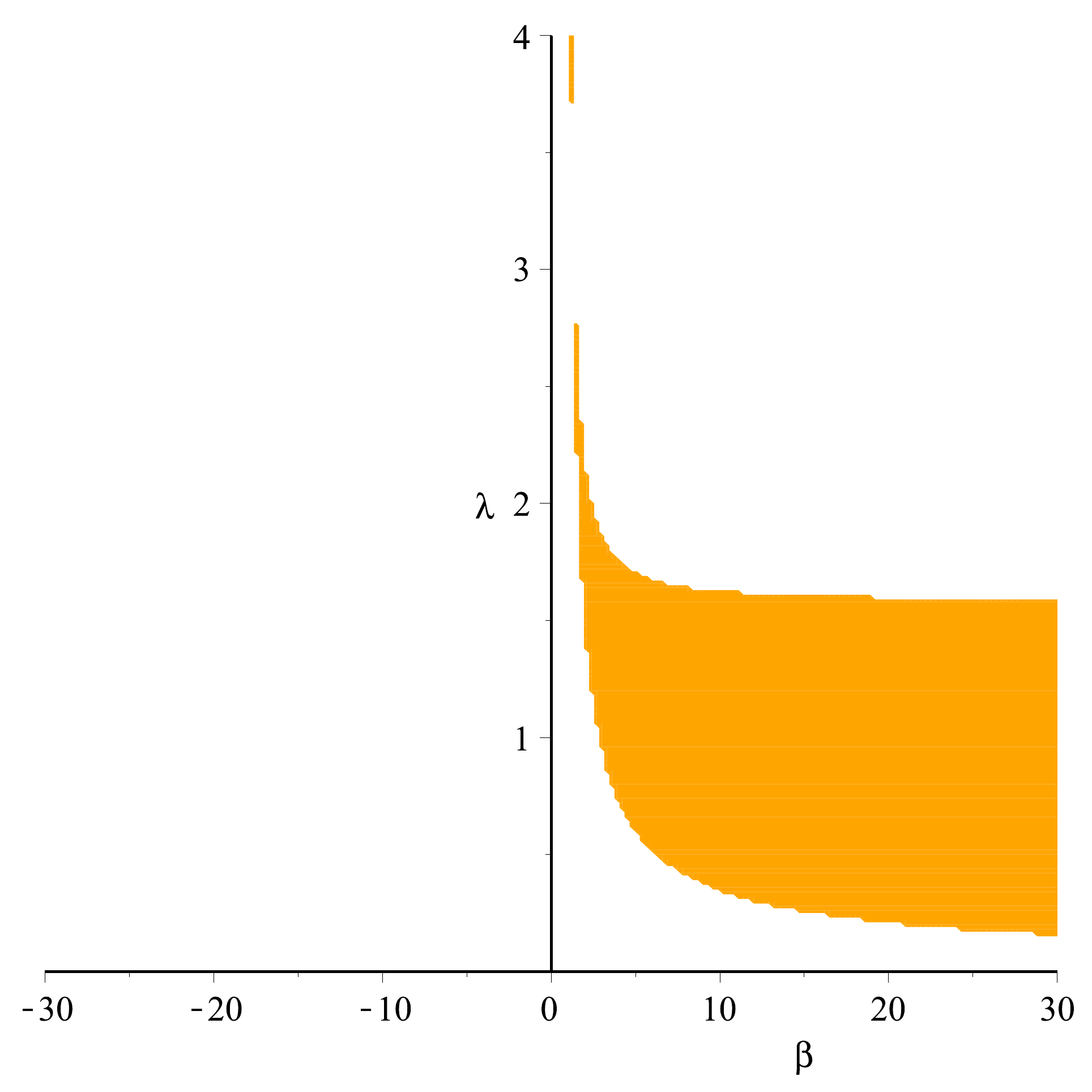}\label{sd5p}}
\qquad
\subfigure[]{%
\includegraphics[width=8cm,height=6cm]{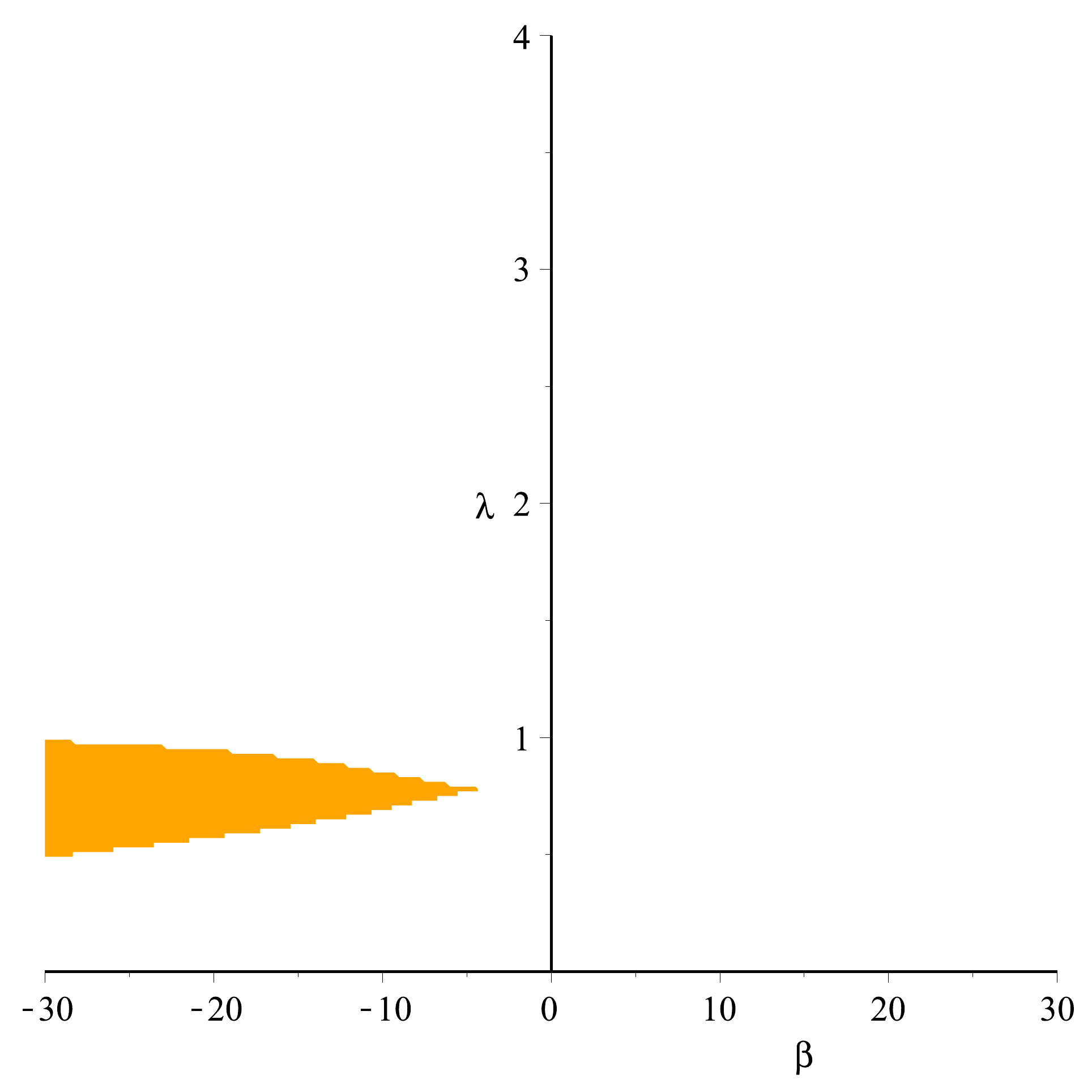}\label{neta}}
\caption{(a). Existence region of point $D_5$. (b). Region for
negativity of one eigenvalue $\eta_+$ with $\theta=-1$.}
\end{figure}
(i) Quintessence field ($\theta=1$):\\
\noindent
Point $D_1$ corresponds to an accelerated matter dominated solution ($\Omega_m=1$) for $\gamma<\frac{2}{3}$.  It is an unstable node for $\frac{1}{1-\beta}<\frac{\gamma}{2}$,
 otherwise it is a saddle point. Scaling solutions $D_2$ and $D_3$ are unstable nodes for $\beta<1$ and are saddle points for $\beta>1$. A set of non isolated
 critical points $D_4$ demands $\lambda=0$ for its existence (i.e., $V(\phi)$ is constant). Since this set is a normally hyperbolic set, so this non isolated set
 of critical points is a late time attractor if $\beta<1$, else it will be saddle. For point $D_5$, since eigenvalues are too complicated to determine  stability
  of the point analytically, we consider the case of dust matter only ($\gamma=1$). We plot a region of existence of the point and region for negativity of one of its eigenvalues
   as shown in figs.\ref{fig11} and \ref{fig12} respectively. We found that these two regions are disjoint, which implies that the eigenvalue must be positive for that point to exist.
    Hence, this point is unstable, for $\gamma=1$. The unstability of point $D_5$ is in contrary with the result in standard GR found in ref \cite{pathak}, where this
     point corresponds to a scaling late time attractor.
\par  In this case, universe evolves from one of these unstable points $D_2$ or $D_3$ and approaches toward the saddle point $D_1$ or $D_5$ and finally settles down towards the attractor set $D_4$.
\begin{center}
\begin{table}[t!]
\caption[crit]{Eigenvalues of critical points in table \ref{Tab
VII}}
\begin{center}
\begin{tabular}{ccccc}
\hline\hline ~~~~Point~~~~&$~~~~~~~E_1~~~~~~~$
&$~~~~~~~E_2~~~~~~~$ &$~~~~~~~~~E_3~~~~~~~~~$&$~~~C_s^2~~~$
\\ \hline\\
$D_{1}$ & $\frac{3}{\beta-1}+\frac{3\gamma}{2}$ &
$\frac{3\gamma}{2}$ & $\frac{3\gamma}{4}$ & undefined(stable
limiting)
 \\ [2ex]
$D_{2}$ & $3\theta(2-\gamma)\,\xi_{+}^2$ & $-\frac{3}{\beta-1}$ &
$-\frac{3}{2(\beta-1)}$ &$1$
\\ [2ex]
$D_{3}$ & $3\theta(2-\gamma)\,\xi_{-}^2$ & $-\frac{3}{\beta-1}$ &
$-\frac{3}{2(\beta-1)}$&$1$
\\ [2ex]
$D_4$ & $-3\gamma$ & $\frac{3}{\beta-1}$ & $0$ &  undefined
\\[2ex]
$D_5$&$\eta_1$&$\eta_+$&$\eta_-$ &Given below \\ \hline \hline
\\\linebreak
\end{tabular} \label{Tab VIII}
\end{center}
For point $D_5$ (quintessence field)
\begin{center}
 $\eta_1=\frac {{\beta}^{2}{\lambda}^{4}-2\,\beta\,{\lambda}^{4}+36\,{
\beta}^{2}{\lambda}^{2}+4\,\beta\,\delta\,{\lambda}^{2}+{\lambda}^{4}-
66\,{\lambda}^{2}\beta-4\,\delta\,{\lambda}^{2}+3\,{\delta}^{2}+30\,{
\lambda}^{2}-27}{32\,{\lambda}^{2} \left( \beta-1 \right) ^{2}}$

$\eta_{\pm}= \frac{1}{{64\,{\lambda}^{2} \left( \beta-1 \right)
^{2}}}\left[192\,{\beta}^{2}{\lambda}^{2}+24\,\beta\,\delta\,{\lambda}^{2}-216\,{\lambda}^{2}\beta-24\,\delta\,{\lambda}^{2}+24\,{\delta}^{2}+24\,{\lambda}^{2}-216\right.$\\$\left.\pm\left(9\,{\beta}^{4}{\lambda}^{8}-36\,{\beta}^{3}{\lambda}^{8}+72\,{\beta}^{4}{\lambda}^{6}+36\,{\beta}^{3}\delta\,{
\lambda}^{6}+54\,{\beta}^{2}{\lambda}^{8}-144\,{\beta}^{3}{\lambda}^{6
}-108\,{\beta}^{2}\delta\,{\lambda}^{6}-36\,\beta\,{\lambda}^{8}-240\,
{\beta}^{3}\delta\,{\lambda}^{4}-10\,{\beta}^{2}{\delta}^{2}{\lambda}^
{4}\right.\right.$\\$\left.\left.+108\,\beta\,\delta\,{\lambda}^{6}+9\,{\lambda}^{8}+432\,{\beta}^{3
}{\lambda}^{4}+624\,{\beta}^{2}\delta\,{\lambda}^{4}+20\,\beta\,{
\delta}^{2}{\lambda}^{4}+144\,\beta\,{\lambda}^{6}-36\,\delta\,{
\lambda}^{6}+72\,{\beta}^{2}{\delta}^{2}{\lambda}^{2}-882\,{\beta}^{2}
{\lambda}^{4}-28\,\beta\,{\delta}^{3}{\lambda}^{2}\right.\right.$\\$\left.\left.-528\,\beta\,\delta\,{\lambda}^{4}-10\,{\delta}^{2}{\lambda}^{4}-72\,{\lambda}^{6}-864\,{
\beta}^{3}{\lambda}^{2}-192\,\beta\,{\delta}^{2}{\lambda}^{2}+468\,
\beta\,{\lambda}^{4}+28\,{\delta}^{3}{\lambda}^{2}+144\,\delta\,{
\lambda}^{4}+1080\,{\beta}^{2}{\lambda}^{2}+828\,\beta\,\delta\,{
\lambda}^{2}\right.\right.$\\$\left.\left.+9\,{\delta}^{4}+120\,{\delta}^{2}{\lambda}^{2}-18\,{
\lambda}^{4}-864\,{\lambda}^{2}\beta-828\,\delta\,{\lambda}^{2}-1296\,
{\beta}^{2}-162\,{\delta}^{2}+648\,{\lambda}^{2}+1296\,\beta+729\right)^{\frac{1}{2}}\right]$\\
$C_s^2=-{\frac
{6\,{\beta}^{2}{\lambda}^{2}+\beta\,\delta\,{\lambda}^{
2}-9\,{\lambda}^{2}\beta-\delta\,{\lambda}^{2}+{\delta}^{2}+3\,{
\lambda}^{2}-9}{3({\lambda}^{2}\beta-{\lambda}^{2}+\delta-3)}}$
\end{center}
\begin{center}
For point $D_5$ (phantom field)\\
$\eta_1=-\frac
{{\beta}^{2}{\lambda}^{4}-2\,\beta\,{\lambda}^{4}-36\,
{\beta}^{2}{\lambda}^{2}+{\lambda}^{4}+66\,\beta\,{\lambda}^{2}+3\,{
\delta}^{2}-30\,{\lambda}^{2}-12\,\delta-27}{{32\lambda}^{2}
\left( \beta-1 \right)^2 }
$\\
$\eta_{\pm}=\frac{1}{{64\,{\lambda}^{2} \left( \beta-1 \right)
^{2}}}\left[192\,{\beta}^{2}{\lambda}^{2}-8\,\beta\,\delta\,{\lambda}^{2
}-216\,\beta\,{\lambda}^{2}+8\,\delta\,{\lambda}^{2}-24\,{\delta}^{2}
+24\,{\lambda}^{2}+96\,\delta+216\right.$\\$\left.\pm4\,\left(9\,{\beta}^{4}{\lambda}^{
8}-36\,{\beta}^{3}{\lambda}^{8}-72\,{\beta}^{4}{\lambda}^{6}-12\,{
\beta}^{3}\delta\,{\lambda}^{6}+54\,{\beta}^{2}{\lambda}^{8}+144\,{
\beta}^{3}{\lambda}^{6}+36\,{\beta}^{2}\delta\,{\lambda}^{6}-36\,
\beta\,{\lambda}^{8}-336\,{\beta}^{3}\delta\,{\lambda}^{4}+86\,{
\beta}^{2}{\delta}^{2}{\lambda}^{4}\right.\right.$\\$\left.\left.-36\,\beta\,\delta\,{\lambda}^{6}
+9\,{\lambda}^{8}+432\,{\beta}^{3}{\lambda}^{4}+648\,{\beta}^{2}
\delta\,{\lambda}^{4}-172\,\beta\,{\delta}^{2}{\lambda}^{4}-144\,
\beta\,{\lambda}^{6}+12\,\delta\,{\lambda}^{6}-72\,{\beta}^{2}{
\delta}^{2}{\lambda}^{2}-882\,{\beta}^{2}{\lambda}^{4}\right.\right.$\\$\left.\left.+52\,\beta\,{
\delta}^{3}{\lambda}^{2}-288\,\beta\,\delta\,{\lambda}^{4}+86\,{
\delta}^{2}{\lambda}^{4}+72\,{\lambda}^{6}+864\,{\beta}^{3}{\lambda}^
{2}+288\,{\beta}^{2}\delta\,{\lambda}^{2}-144\,\beta\,{\delta}^{2}{
\lambda}^{2}+468\,\beta\,{\lambda}^{4}-52\,{\delta}^{3}{\lambda}^{2}\right.\right.$\\$\left.\left.-24\,\delta\,{\lambda}^{4}-1080\,{\beta}^{2}{\lambda}^{2}-1044\,\beta
\,\delta\,{\lambda}^{2}+9\,{\delta}^{4}+216\,{\delta}^{2}{\lambda}^{2}
-18\,{\lambda}^{4}+864\,\beta\,{\lambda}^{2}-72\,{\delta}^{3}+756\,
\delta\,{\lambda}^{2}-1296\,{\beta}^{2}\right.\right.$\\$\left.\left.-18\,{\delta}^{2}-648\,{
\lambda}^{2}+1296\,\beta+648\,\delta+729\right)^{\frac{1}{2}}\right]$\\
$C_s^2=-{\frac
{6\,{\beta}^{2}{\lambda}^{2}-\beta\,\delta\,{\lambda}^{
2}-9\,{\lambda}^{2}\beta+\delta\,{\lambda}^{2}-{\delta}^{2}+3\,{
\lambda}^{2}+6\,\delta+9}{3({\lambda}^{2}\beta-{\lambda}^{2}-\delta+3)}}$

\end{center}
\end{table}
\end{center}

\par The behaviour of deceleration parameter $q$ for the case of $\Gamma=1$ is given in fig.\ref{decpq}. The universe undergoes transition from
decelerated phase to an accelerated phase around $N=-0.44$ (equivalent to a redshift $0.55$) which match with the observation \cite{Liang}. Finally
 the universe settles down to an accelerated expansion ($q=-1$).  Also crossing of phantom divide is possible as shown in fig.\ref{weffpq} for quintessence field. A similar behaviour can be observed for the case of $\Gamma\neq 1$ also. In this case also the crossing of phantom divide line can be understood analytically as in case of interaction {\bf A}.\\\linebreak
(ii) Phantom field ($\theta=-1$):\\
Critical points $D_2,\, D_3$ do not exist for phantom field. Critical point $D_1$ corresponds to an accelerated matter dominated phase for $\gamma<\frac{2}{3}$. It is unstable node if $\frac{1}{1-\beta}<\frac{\gamma}{2}$, else it is a saddle point. A set of non isolated critical points $D_4$ corresponds to an accelerated solution. Again, since its Jacobian matrix contains two negative and one zero eigenvalues, further investigation is required. We plot the projection of the system on $x-y$ plane and from (fig.\ref{sppe}) we observe that point $(0,1)$ which lies on the set $D_4$ is not stable. Therefore, the set $D_4$ is not a stable set. For critical point $D_5$ since it is too complicated to determine its stability, we focus only in the case of dust matter $(\gamma=1)$. As before, we plot the existence region of a point (fig.\ref{sd5p}) and the region of negativity of one of its eigenvalue (fig.\ref{neta}). It is observed that these two regions are disjoint which implies that the point is not stable.
\begin{figure}
\centering
\subfigure[]{%
\includegraphics[width=7cm,height=5cm]{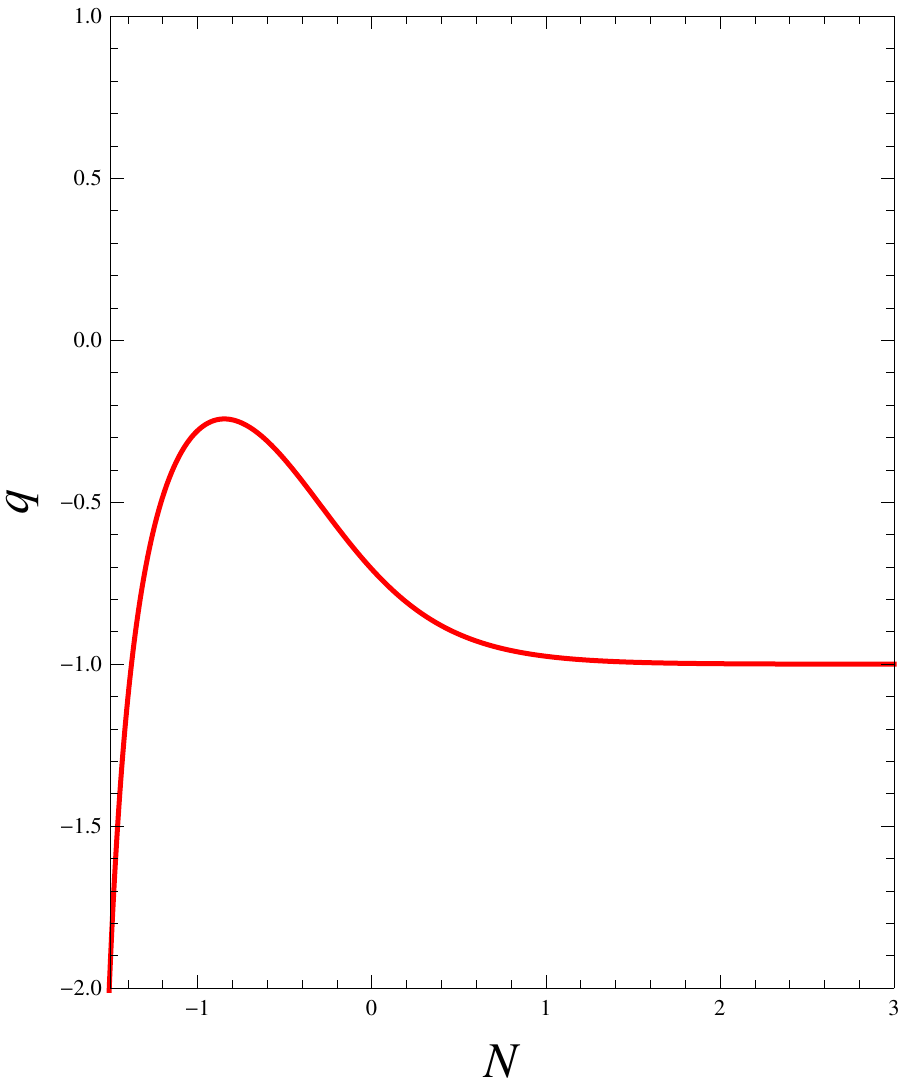}\label{decpp}}
\qquad
\subfigure[]{%
\includegraphics[width=7cm,height=5cm]{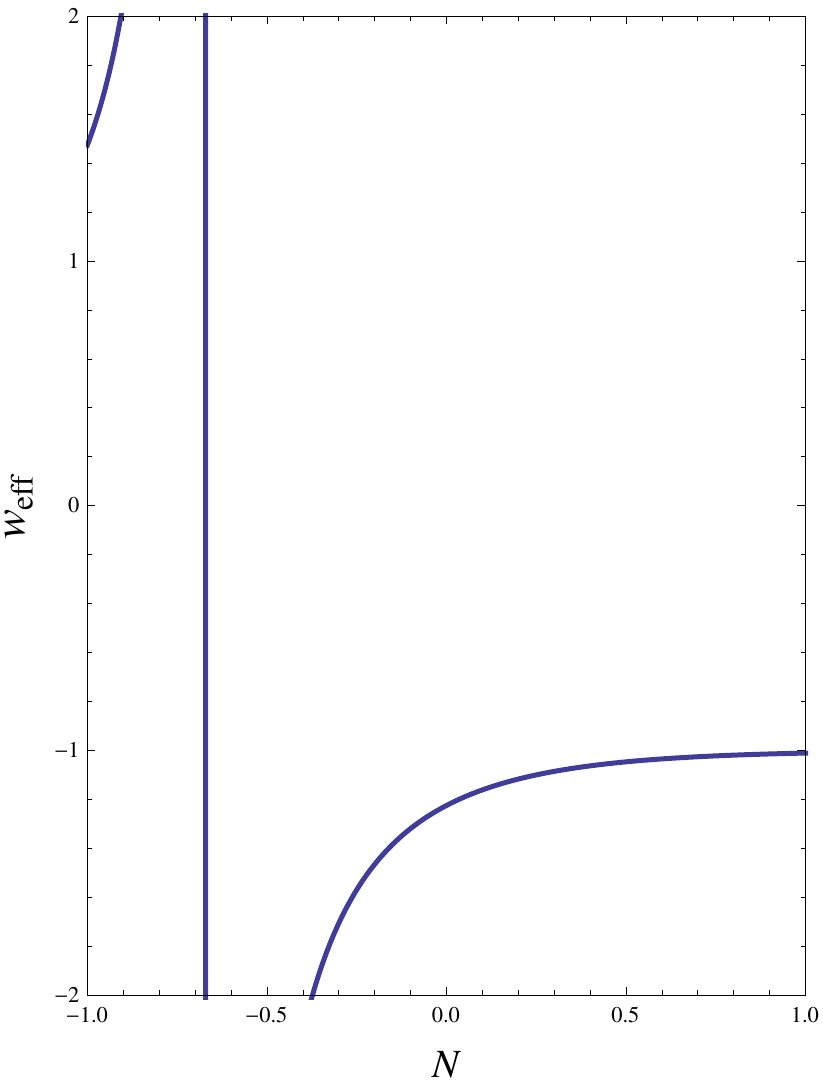}\label{weffpp}}
\caption{ (a). Plot of $q$ versus $N$. (b). Plot of $\omega_{\rm eff}$ \textit{vs N}  for  $\theta=-1$, $\Gamma=1$ with $\beta=-0.7$, $\gamma=1$.}
\end{figure}

\par In contrast to interaction \textbf{A}, we could not extract any late time accelerated scaling attractors in this case. However, for $D_5$, $\gamma \neq 1$ may give some interesting solution.
\par The behaviour of deceleration parameter $q$ for the case of $\Gamma=1$ is given in fig.\ref{decpp}. The universe is always in accelerated phase in this case.
 Also crossing of phantom divide is not possible as shown in fig.\ref{weffpp}. As in case of interaction \textbf{A}, there is a breakdown due to the effective behaviour and no pathology is associated with the model.

\section{\textbf{Concluding remarks}}\label{sec5:conc}
\noindent The present work deals with the dynamical system analysis of interacting DE in flat DGP model. The source of DE is taken to be scalar field (quintessence/phantom)
 and two specific interactions are considered for stability of critical points. Further the stability of critical points are examined for two categories of potentials (exponential and non
 exponential). The potential is classified in two categories as investigation for these two categories gives complete possible set of choices. For each interaction, we have studied four sub-cases. Performing a detailed stability analysis for each category of potentials for two types of scalar fields we have extracted late time attractors for each interaction along with all important cosmological parameters. Finally in order to predict ultimate fate of evolution, we have also examined classical stability of each critical point for different cases.
In what follows we summarise our main results.
\par While in interaction $\mathbf{A}$, there is no matter dominated phase for late time attractors, we do get matter dominated phase for interaction $\mathbf{B}$.
\par For interaction \textbf{A}  and non exponential potential we get late
 time accelerated attractor $A_4$ for quintessence field only. The phantom field in this case does not give any physically interesting result.
  However for exponential potential, phantom field yields a late time accelerated scaling solution $B_5$. Further, this solution is also classically stable.
  This is very interesting case from cosmological point of view. Moreover, exponential potential also admits in this case late time accelerated attractor
  $B_6$ for quintessence field. This result is in contrast to standard cosmology where interacting quintessence only admits late time accelerated scaling solution.
\par For quintessence field, the interaction $\mathbf{B}$ gives similar results for both categories of potentials. Phantom field does not give any interesting cosmological solution for both categories of potentials in interaction $\mathbf{B}$. It may be noted that due to complicated calculations in the phantom field for the exponential potential, we have examined the point $D_5$ for $\gamma = 1$ only.  So the possibility of having late time accelerated scaling attractor in this case cannot be ruled out for $\gamma \neq 1$. It will be interesting to choose $\gamma$ in such a way that it gives late time accelerated scaling solution in this interaction also. We leave it for our future work.\\\linebreak
\textbf{Acknowledgement}\\
The authors wish to thank the anonymous referees for helpful
suggestions which lead to further improvement of this work.


\begin{thebibliography}{99}
\bibitem{Perlmutter}
S. J. Perlmutter $ et\, al$, Astrophys. J. \textbf{517}, 565 (1999).
\bibitem{Spergel}
D. N. Spergel \textit{et al}, Astrophys J. Suppl. \textbf{148},175 (2003).
\bibitem{Riess}
A. G. Riess \textit{et al}, Astrophys. J. \textbf{607}, 665 (2004).
\bibitem{Ade}
P. A. R. Ade \textit{et al} (Planck Collaboration) Astron Astrophys $\mathbf{571}$, A$16$ (2014).
\bibitem{Carroll}
S.M. Carroll, Living Rev.Rel. \textbf{4}, 1 (2001).[arXiv:astro-ph/0004075].
\bibitem{Peebles}
P. Peebles, P. Ratra, Rev.Mod.Phys. \textbf{75} 559-606 (2003). [arXiv:astro-ph/0207347]
\bibitem{Lyth}
A. R. Liddle, D. H. Lyth, Cosmological inflation and Large Scale Structure. Cambridge University Press, Cambridge,
England, (2003).
\bibitem{Matos}
J. Magana, T. Matos, J. Phys. Conf. Ser. \textbf{378}, 012012 (2012).
\bibitem{Linder}
E.V. Linder, Gen. Rel. Grav. \textbf{40}, 329 (2008). [arxiv:704.2064]
\bibitem{Armendariz}
C. Armendariz-Picon, V.F. Mukhanov, P.J. Steinhardt, Phys. Rev. D
\textbf{63} 103510 (2001). [arXiv:astro-ph/0006373]
\bibitem{Caldwell}
R.R. Caldwell, Phys. Lett. B \textbf{545} 23-29 (2002). [arXiv:astro-ph/9908168]
\bibitem{cope}
E.J. Copeland, M. Sami, S. Tsujikawa, Int. J. Modern Phys. D \textbf{15}, 1753  (2006).
\bibitem{Bamba}
K. Bamba, S. Capozziello, S. Nojiri, S. D. Odintsov, Astrophys. Space Sci. $\mathbf{342}$ 155-228 (2012).
\bibitem{Kalara}
J. Ellis, S. Kalara, K. A. Olive and C. Wetterich, Phys. Lett. B $\mathbf{228}$, 264 (1989).
\bibitem{Wetterich}
C. Wetterich, Astron. Astrophys. $\mathbf{301}$, 321 (1995).
\bibitem{Amendola}
L. Amendola, Phys. Rev. D $\mathbf{60}$ 043501 (1999).  [arxiv:astro-ph/9904120]
\bibitem{Guo}
Z. K. Guo and Y. Z. Zhang, Phys. Rev. D $\mathbf{71}$, 023501 (2005).[arXiv:astro-ph/0411524]
\bibitem{Nojiri}
S. Nojiri, S. D. Odintsov and S. Tsujikawa,  Phys. Rev. D $\mathbf{71}$, 063004 (2005). [arXiv:hep-th/0501025].
\bibitem{Marteens}
R. Maartens, Living Rev. Relativ. {\bf 7}, 7 (2004).
\bibitem{dgp1}
G. R. Dvali, G. Gabadadze, M. Porrati, Phys.Lett. B \textbf{485},208 (2000).
\bibitem{dgp2}
C. Deffayet, Phys Lett. B \textbf{502}, 199 (2001).
\bibitem{JDutta1}
  J.~Dutta, S.~Chakraborty and M.~Ansari,
  Int.\ J.\ Theor.\ Phys.\  {\bf 49}, 2680 (2010)
  [arXiv:1006.2206 [gr-qc]].
\bibitem{JDutta2}
  J.~Dutta, S.~Chakraborty and M.~Ansari,
  Mod.\ Phys.\ Lett.\ A {\bf 25}, 3069 (2010)
  [arXiv:1005.5321 [gr-qc]].
\bibitem{JDutta3}
  J.~Dutta and S.~Chakraborty,
  Int.\ J.\ Theor.\ Phys.\  {\bf 50}, 2383 (2011)
  [arXiv:1006.2210 [gr-qc]].
\bibitem{Lazkoz}
R. Lazkoz, R. Marteens, E. Majerotto, Phys. Rev. D \textbf{74}, 083510 (2006). [arxiv:astro-ph/ 0605701].
\bibitem{Lue}
A. Lue and G.D Starkman, Phys. Rev. D\textbf{ 70}, 101501 (2004) [arXiv:astro-ph/0408246].
\bibitem{Chimento}
P. L. Chimento, R. Lazkoz, R. Maartens, I. Quiros,  JCAP{\bf 09}, 004
(2006). [arxiv:astro-ph/ 0605450].
\bibitem{Lopez}
M. Bouhmadi-Lopez and R. Lazkoz Phys. Lett. B \textbf{654}, 51 (2007). [arXiv:astro-ph/0706.3896].
\bibitem{Zhang}
H. Zhang and  Z. H. Zhu, Phys. Rev. D {\bf 75}, 023510 (2007).
\bibitem{Wu}
X. Wu, R. G.  Cai, and Z. H. Zhu Phys. Rev. D \textbf{77}, 043502
(2008).
\bibitem{Ellis}
J. Wainwright, G. F. R. Ellis, \textit{Dynamical Systems in
Cosmology}. (Cambridge University Press, 1997).
\bibitem{Coley}
A. A. Coley, \textit{Dynamical systems and cosmology}. (Kluwer
Academic Publishers, Dordrecht Boston London, 2003).
\bibitem{cg}
C. G. Boehmer, G. Caldera-Calbral, R. Lazkoz, R. Maartens,
Phys. Rev. D \textbf{78} 023505 (2008). [arxiv:gr-gc 0801.1565]
\bibitem{Cai}
Z. K. Guo, R. G. Cai, Y. Z. Zhang, JCAP{\bf 05}, 002 (2005).
[arxiv:astro-ph/0412624]
\bibitem{Mahata:2015nga}
N.~Mahata and S.~Chakraborty,
  Mod.\ Phys.\ Lett.\ A {\bf 30}, no. 27, 1550134 (2015).
\bibitem{Biswas:2015cva}
  S.~K.~Biswas and S.~Chakraborty,
  Int.\ J.\ Mod.\ Phys.\ D {\bf 24}, no. 07, 1550046 (2015)
\bibitem{Biswas:2015zka}
  S.~K.~Biswas and S.~Chakraborty,
  Gen.\ Rel.\ Grav.\  {\bf 47}, 22 (2015)
\bibitem{pathak}
M. Shahalam, S. D. Pathak, M. M. Verma, M. Y. Khlopov and R.
Myrzakulov, Euro. Phys. J. C75 \textbf{8}, 395 (2015).
[arxiv:gr-gc/1503.08712]
\bibitem{Liddle}
E.~J. Copeland, A. R.  Liddle and D. Wands, Phys. Rev. D
$\mathbf{57}$, 4686-4690 (1998) [gr-qc/9711068]
\bibitem{Scherer}
A. R. Liddle and R. J. Scherrer, Phys. Rev. D $\mathbf{59}, 023509\, (1999) $[arXiv:astro-ph/9809272]
\bibitem{Sujikawa}
S. Tsujikawa and M. Sami, Phys. Lett. B $\mathbf{603},\,113\, (2004)$ [arXiv:hep-th/0409212].
\bibitem{Quiros}
I. Quiros, R. Garcia-Salcedo, T. Matos, C.  Moreno,
Phys. Lett. B\textbf{670}, 259-265 (2009) [arxiv:gr-gc 0802.3362]
\bibitem{Leyva}
Y.  Leyva, D. Gonzalez, T.  Gonzalez, T. Matos, I. Quiros, Phys. Rev. D \textbf{ 80}, 044026 (2009). [arXiv:gr-gc 0909.0281].
\bibitem{Nozari}
K. Nozari, F. Rajabi, K. Asadi, Class. Quantum. Grav. $\mathbf{29}$, 175002 (2012). [arxiv:gr-gc 1208.1666]
\bibitem{Miller}
 A. D. Miller \textit{et.al}., Astrophys. J. Lett. \textbf{524}, L1 (1999).
\bibitem{Liang}
N. Liang, Z. H. Zhu, Research in Astron. Astrophys 497-506 (2011).
\bibitem{Chao}
L. Jie-Chao \textit{et al}, Chin. Phys. Lett Vol No. \textbf{2}, 802 (2008).
\bibitem{Wiggins}
S. Wiggins, \textit{Introduction to Applied Nonlinear Dynamical Systems and Chaos}. (Springer, New York Heidelberg Berlin, 1990).
\bibitem{Strogatz}
S. H. Strogatz, \textit{Nonlinear Dynamics and Chaos: With
Applications to Physics, Biology Chemistry and Engineering}
(Westview Press, Boulder, 2001).
\bibitem{Perko}
L. Perko, \textit{Differential Equations and Dynamical Systems}. (SpringerVerlag, 1991).
\bibitem{Aulbach}
B. Aulbach, \textit{Continuous and Discrete Dynamics near Manifolds of Equilibria}. (Lecture Notes in Mathematics No. 1058, Springer, 1984).
\bibitem{Nandan}
N. Roy, N. Banerjee, Euro. Phys. J. Plus. \textbf{129}, 162 (2014).
\bibitem{Jibitesh}
J. Dutta, ~H. Zonunmuiwah, Euro. Phys. J. Plus. $\mathbf{130},
221\,(2015).$
\bibitem{Billiyard}
A. P. Billiyard, A. A. Coley, Phys. Rev D\textbf{61}, 083503, 2000.
[arxiv:astro-ph/9908224]
\bibitem{Amen}
L. Amendola and S. Tsujikawa, Dark Energy Theory and Observations, Cambridge University Press, Cambridge UK, (2010).
\bibitem{NandanRoy}
N. Roy, N. Banerjee, Annals Phys.\ {\bf 356}, 452 (2015).
\bibitem{Steinhardt}
I. Zlatev, L. M. Wang, P. J. Steinhardt, Phys. Rev. Lett.$
\mathbf{82},\, 896\, (1999)$[arxiv:astro-ph/9807002]
\bibitem{Billy_thesis}
A.P. Billyard, The Asymptotic Behaviour of Cosmological Models Containing Matter and Scalar Fields (PhD thesis, Dalhousie University, 1999).
\bibitem{Lucchin}
F. Lucchin and S. Matarrese, Phys. Rev. D {\bf 32}, 1316 (1985).
\bibitem{Wett}
C. Wetterich, Nucl. Phys. B {\bf 302}, 668 (1988)
\bibitem{Wands}
D. Wands, E. J. Copeland and A. R. Liddle, Ann. N. Y. Acad. Sci. {\bf 688}, 647 (1993).
\bibitem{Piazza}
N. Piazza,~S. Tsujikawa, JCAP \textbf{0407}, 004 (2004).
\bibitem{Mahata}
N. Mahata,~S.  Chakraborty, Gen. Rel. Grav. \textbf{46},1721
(2014).
\end{thebibliography}
\end{document}